\documentclass[%
prl,
 amsmath,amssymb,
reprint,%
longbibliography]{revtex4-1}

\usepackage{graphicx}
\usepackage{dcolumn}
\usepackage{bm}
\usepackage{SIunits}
\usepackage{float}
\usepackage{color}

\usepackage[normalem]{ulem}

\newcommand{\omegap}{\omega_\text{p}}
\newcommand{\omegac}{\omega_\text{c}}
\newcommand{\omegam}{\omega_{\textrm{m}}}
\newcommand{\omegak}{\omega_{k}}
\newcommand{\omegaL}{\omega_{\textrm{L}}}
\newcommand{\omegaFSR}{\omega_{\textrm{FSR}}}

\newcommand{\opd}[2]{\mbox{$\hat{#1}_{#2}^{\dagger}$}}  
\newcommand{\op}[2]{\mbox{$\hat{#1}_{#2}$}}

\newcommand{\kappae}{\kappa_{\textrm{e}}}
\newcommand{\kappai}{\kappa_{\textrm{i}}}

\newcommand{\aout}{a_{\textrm{out}}}
\newcommand{\ain}{a_{\textrm{in}}}
\newcommand{\aopd}{\hat{a}^\dagger}
\newcommand{\deltak}{\delta_{k}}
\newcommand{\aop}{\hat{a}}

\newcommand{\bopd}{\hat{b}^\dagger}
\newcommand{\bop}{\hat{b}}
\newcommand{\ckop}{\hat{c}_{k}}

\newcommand{\gammai}{\gamma_{\textrm{i}}}

\newcommand{\gammaki}{\gamma_{k,\textrm{i}}}

\newcommand{\gammak}{\gamma_{k}}
 
\newcommand{\gammakS}{\gamma_{k,\textrm{S}}}
\newcommand{\gammaOM}{\gamma_{\textrm{OM}}}
\newcommand{\gammae}{\gamma_{\textrm{e}}}
\newcommand{\nintrinsic}{n_{\textrm{i}}} 
\newcommand{\nc}{n_{\textrm{c}}}
\newcommand{\np}{n_{\textrm{p}}}
\newcommand{\gammap}{\gamma_{\textrm{p}}}
\newcommand{\nb}{n_{\textrm{b}}}
\newcommand{\vg}{v_{\textrm{g}}}
\newcommand{\etas}{\eta_{\textrm{s}}}
\newcommand{\etaf}{\eta_{\textrm{f}}}
\newcommand{\etac}{\eta_{\textrm{c}}}

\newcommand{\vdet}{{v_{\textrm{det}}}}
\newcommand{\alphaL}{\alpha_{\textrm{L}}}
\newcommand{\alphap}{\alpha_{\textrm{p}}}
\newcommand{\alphaM}{\alpha_{\textrm{m}}}

\newcommand{\Kout}{\check{K}_{\textrm{out}}}
\newcommand{\Kin}{\check{K}_{\textrm{in}}}

\newcommand{\Cwg}{C_{\textrm{wg}}}

\begin{document}


\title[A single-mode phonon waveguide]{A single-mode phononic wire}

\author{Rishi N. Patel}
\email{rishipat@stanford.edu}
\author{Zhaoyou Wang}
\author{Wentao Jiang}
\author{Christopher J. Sarabalis}
\author{Jeff T. Hill}
\author{Amir H. Safavi-Naeini}%
\email{safavi@stanford.edu}
\affiliation{%
Department of Applied Physics and Ginzton Laboratory, Stanford University
}%

\date{\today}
\begin{abstract}

Photons and electrons transmit information to form complex systems and networks. Phonons on the other hand, the quanta of mechanical motion, are often considered only as carriers of thermal energy. Nonetheless, their flow can also be molded  in fabricated nanoscale circuits. We design and experimentally demonstrate wires for phonons that  transmit information with little loss or scattering across a chip. By patterning the surface of a silicon chip, we completely eliminate all but one channel of phonon conduction. We observe the emergence of low-loss standing waves in millimeter long phononic wires that we address and cool optically. Coherent transport and strong optical coupling to a phononic wire enables new phononic technologies to manipulate information and energy on a chip.
\end{abstract}

\maketitle

Engineered structures guide waves of light, sound, and electrons -- transmitting information and energy across distances from microns to kilometers. The most prominent examples are the optical waveguides that form the physical basis of the internet, integrated photonic technologies~\cite{Soref2006}, nonlinear optics~\cite{boyd2003nonlinear}, and the emerging quantum internet~\cite{Kimble2008b,Lodahl2017a,Reitz2013}. The mechanical analog of these electromagnetic structures are \emph{phononic wires} or \emph{waveguides} that would coherently transport mechanical energy and information in highly confined waves. Currently established approaches for guiding mechanical waves over many wavelengths use surface or bulk acoustic waves that are unconfined in the direction transverse to propagation~\cite{Gustafsson2012,OConnell2010,Chueaao1511,Han2016}. Attempting to confine these waves leads to large scattering and diffractive losses and prevents the development of complex phononic circuitry and confinement-enhanced interactions. Additionally, highly confined mechanical waveguides are difficult to excite electromagnetically by direct capacitive or piezoelectric coupling due to the vanishing capacitance of small transducers. These challenges have prevented the emergence of phonon networks such as those recently proposed to distribute quantum information on a chip~\cite{Habraken2012,Vermersch2017,Cirac1996}.
\begin{figure*}[t]
\includegraphics[scale=0.98]{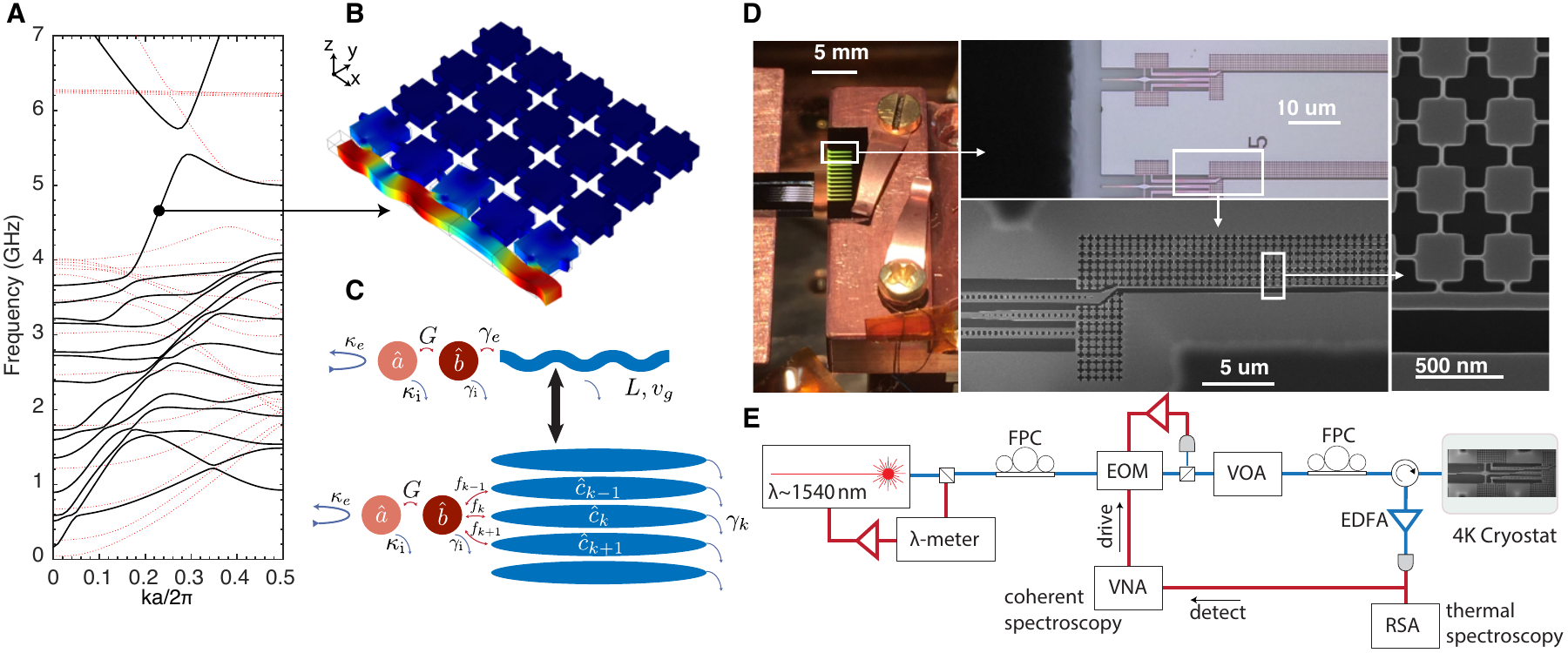}
\caption{\label{fig:Overview} \textbf{System Concept and Experiment Overview} \textbf{(A)} Band diagram for the single-mode phonon waveguide. The solid black (dotted-red) lines show the z-symmetric (antisymmetric) bands respectively. \textbf{(B)} Simulated displacement field for a longitudinal edge-mode in the bandgap. \textbf{(C)} Concept of the readout mechanism, showing an optical mode $\aop$, coupled to a localized mechanical resonance $\bop$. The localized mechanical mode can be used to readout a terminated single-mode waveguide. On the bottom is shown an equivalent picture, in which we treat each waveguide mode as a standing-wave coupled at a rate $f_{k}$. \textbf{(D)} Experimental realization of the device, showing the cryostat setup (along the direction of propagation, the $2~\milli\meter$ on-chip waveguides can be resolved with the unaided eye), a laser confocal image, and two scanning electron micrographs.\textbf{(E)} Simplified diagram of the optical setup. A laser is used to probe an optomechanical cavity in reflection. EOM: Electro-optic modulator, VOA: Variable optical attenuator, EDFA: Erbium-doped fiber amplifier, VNA: Vector network analyzer, RSA: Realtime spectrum analyzer, FPC: Fiber polarization controller.} 
\end{figure*}

We report the first demonstration of a single-mode phononic wire.  Coupling the wire to a localized optomechanical transducer allows us to excite and detect phonons, and to observe their low-loss, coherent transport across a chip. We strongly couple the phononic wire to the light field by detuned laser driving of the optomechanical transducer, resulting in sympathetic cooling of the wire to a temperature below that of the surrounding cryogenic environment.

Phonon conduction in a quantum wire can occur through four independent channels~\cite{Schwab2000}. These channels correspond to two flexural excitations, a dilatational, and a torsional wave. All channels persist in the long-wavelength limit down to zero frequency. The existence of multiple polarizations complicates phonon transport over long distances since small amounts of disorder lead to interchannel scattering and a rapid loss of coherence. This scattering makes the response of the system sensitive to details reminiscent of quantum chaos in mesoscopic samples~\cite{Marcus1992,Baranger1993}. Moreover, interchannel scattering is enhanced by going to highly confining structures as needed to push higher-order phonon conduction channels into cut-off.

We overcome these challenges by nanoscale patterning of phononic materials. Phononic crystals use periodicity to significantly modify the mode spectrum for mechanical waves~\cite{Maldovan2013,Hatanaka2014a}. The most dramatic effect is the elimination of phonons of all polarizations for a given range of frequencies by formation of a full phononic bandgap \cite{Chan2012,Safavi-Naeini2014a}. We induce a line defect on a phononic crystal possessing a full bandgap to generate a structure where only a single propagating mode of group velocity $v_{g} \approx 6800~\meter\per\second$ is allowed within a large range of frequencies (Fig.~\ref{fig:Overview}A). The waveguide is fabricated in a suspended film of silicon with thickness $220~\text{nm}$ using a combination of e-beam lithography, and dry and wet etching~\cite{Patel2017}. 

To characterize the properties of a long waveguide, we terminate one end with a reflective boundary, and place a cavity-optomechanical transducer that supports localized optical and mechanical modes at the opposite end. As shown in Fig.~\ref{fig:Overview}C the waveguide of length $L$, which supports traveling waves of group velocity $\vg$, can be analyzed by considering standing-wave modes of an extended cavity. Each extended mode of the wire is coupled to the localized mechanical resonance of the transducer at a rate $f_{k}$.
In the frequency domain, we describe the interaction between the extended modes and the localized mechanical mode as
\begin{equation}
\label{eqn:wgint}
\hat{H}_{\textrm{wg,int}}/\hbar = \sum_{k} (f_{k}\ckop^{\dagger}\bop + h.c.),
\end{equation}
where we take $f_{k}$ to be real. Operators $\ckop$  and $\bop$ annihilate phonons in mode $k$ of the wire and the localized mode, respectively. Under the effect of a red-detuned laser drive, the optomechanical transducer effectively couples the mechanical resonance to the optical field leading to a total Hamiltonian
\begin{equation}
\hat{H}_{}/\hbar = \Delta \opd{a}{} \op{a}{}+\omegam \opd{b}{} \op{b}{}+G(\opd{a}{}\op{b}{} + h.c.) + \hat{H}_{\textrm{wg,int}}/\hbar, \label{eqn:total_ham}
\end{equation}
where $G = g_{0}\sqrt{\nc}$ is the laser-enhanced optomechanical coupling rate for $\nc$ pump photons, $\aop$ is the cavity photon annihilation operator, and $\Delta \equiv \omegac - \omegaL {\approx \omegam}$ is the laser-cavity detuning. 

\begin{figure}[h]
\includegraphics[scale=1]{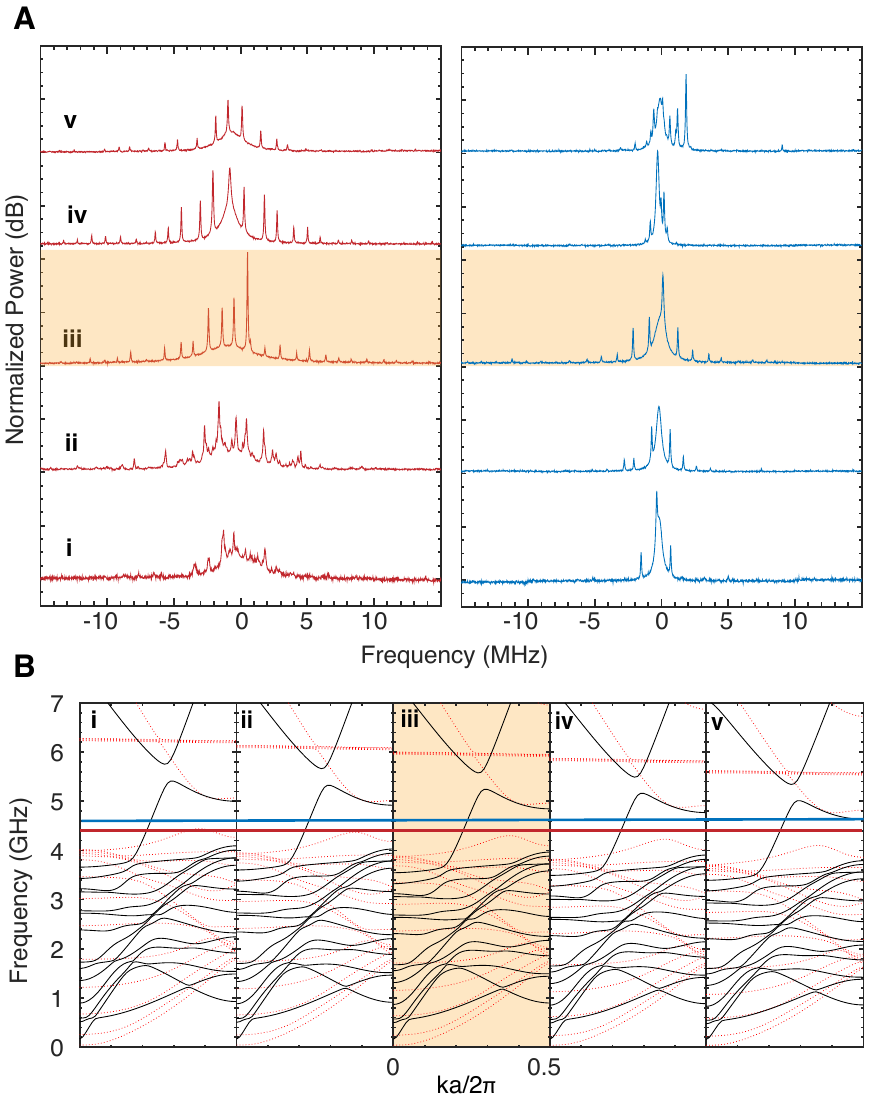} 
\caption{\label{fig:structure-tuning}\textbf{Probing the Single Mode Regime via Structural Tuning} \textbf{(A)} Experimentally measured thermal power spectra of the waveguide-cavity device as the lattice constant is changed across five different devices, increasing from bottom to top. The left (right) sides show the spectra from localized readout mode 1 (2) respectively. $\omega_{\textrm{m,1}}/2\pi \approx 4.4~\giga\hertz$ and $\omega_{\textrm{m,2}}/2\pi \approx 4.6~\giga\hertz$. The labeled lattice constant sweep values are $a = 450~\nano\meter(\textrm{i}),457~\nano\meter(\textrm{ii}),464~\nano\meter(\textrm{iii}),472~\nano\meter(\textrm{iv}),486~\nano\meter(\textrm{v})$. The highlight corresponds to both localized modes in the single mode regime. \textbf{(B)} Band diagram of the waveguide for each lattice constant. The red and blue horizontal lines denote the frequencies of the two transducer modes.}
\end{figure}

A simplified diagram of the experimental setup used to perform low-temperature ($T = 11~\kelvin$) coherent and thermal spectroscopy is shown in Fig.~\ref{fig:Overview}E. In coherent spectroscopy, in addition to the strong pump described above, we generate a weak probe tone that is swept across the cavity resonance. The probe experiences a large phase shift at pump-probe detunings near the mechanical resonances of the structure. This is a generalization of electromagnetically induced transparency (EIT) for optomechanical systems~\cite{Weis2010,Safavi-Naeini2010h,Massel2012,Safavi-Naeini2014}. Separately, the incoherent spectrum of the thermal phonons scattered into a sideband of the optical pump field are measured by analyzing the power spectral density of the detected light field. An optomechanical transducer uncoupled from a phononic wire would have a thermal spectrum that is a Lorentzian line centered at the mechanical frequency arising from losses into a nearly Markovian thermal bath. In contrast, the phononic wire acts as a strongly non-Markovian bath and modifies this detected spectrum leading to the emergence of multiple peaks as shown in Fig. ~\ref{fig:structure-tuning}A.

Thermal spectroscopy allows us to experimentally probe the band structure of phononic wires. We fabricate an array of transducer-coupled 3-millimeter-long phononic wires and measure them in the cryostat. Since each localized mechanical mode of the transducer probes the waveguide spectrum over a narrow range of at most a few megahertz, we fabricate multiple optomechanically coupled phononic wires and probe them independently. The results of thermal spectroscopy performed on the two separate transducer modes with frequencies separated by roughly $200~\text{MHz}$ are presented by the red and blue solid lines in Fig.~\ref{fig:structure-tuning}.  During fabrication, we sweep the phononic wire lattice constant across devices over an $8\%$ range, effectively sweeping the single-mode region of the wires over the frequencies of the two localized readout modes (Fig.~\ref{fig:structure-tuning}B). For each device, we record the normalized thermal spectrum as shown in Fig.~\ref{fig:structure-tuning}A. The regime of single-mode operation becomes immediately apparent through the emergence of regularly-spaced peaks on top of the broad thermal noise pedestal of the transducer signal (Fig.~\ref{fig:structure-tuning}A,iii). The thermal spectra corresponding to multi-band propagation are qualitatively different --  they do not have a well-defined free spectral range (FSR) due to mixing between different phonon propagation channels (Fig.~\ref{fig:structure-tuning}A, i and ii). In the single-mode case, by measuring the peak separations, we find a mode spacing $\overline{\Delta \omegak}/2\pi=1.1\pm0.2~\mega\hertz$, agreeing with the expected FSR of $v_{\textrm{g}}/2L = 1.14~\mega\hertz$. This corresponds to a round-trip time of approximately $900~\text{ns}$.

\begin{figure*}
\includegraphics[scale=1.0]{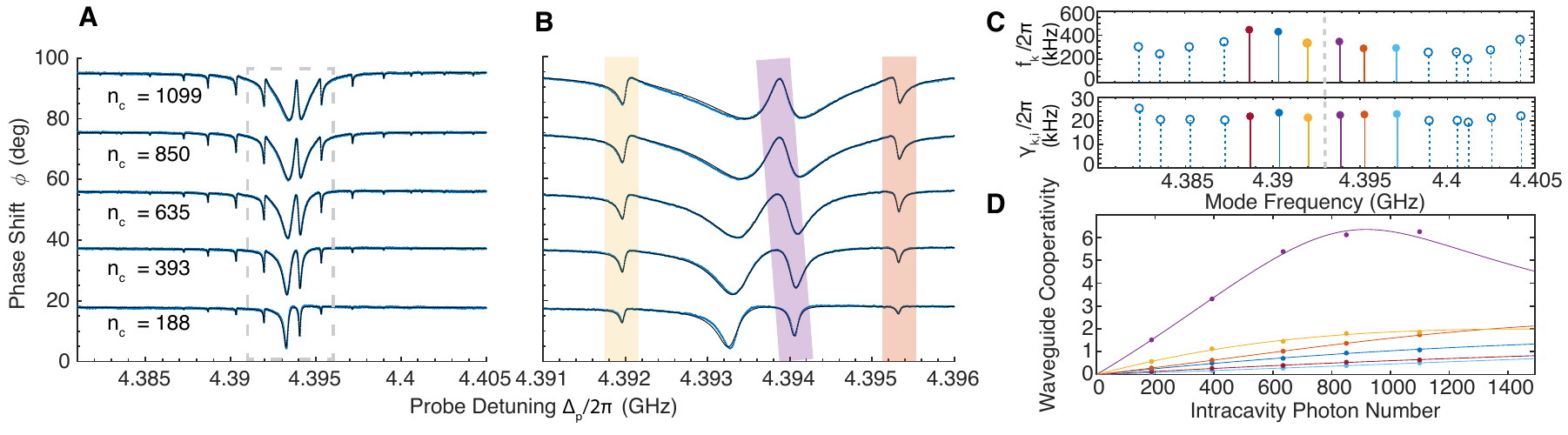}
\caption{\label{EITdata} \textbf{Coherent Linear Spectroscopy Results} \textbf{(A)} Phase response of a weak coherent probe tone as a function of detuning from the pump beam. Several coupled waveguide resonances are seen. The central dip broadens with laser power, due to sideband cooling of the localized mechanical mode. The black line is a fit to a theoretical model, described in the main text. \textbf{(B)} A zoom in of data from (A), showing three waveguide modes (highlighted) and the localized mode near resonance. The localized mode is strongly mixed with the waveguide mode highlighted in purple. \textbf{(C)} Model parameters estimated from dataset in (A), showing the waveguide-coupling rates $f_{k}$ (top) and the intrinsic loss rates $\gamma_{k,\textrm{i}}$ (bottom) vs. waveguide mode frequency. The solid lines correspond to lines in (D). The gray dashed line shows the approximate frequency of the localized mechanical resonance. \textbf{(D)} Waveguide cooperativities inferred from model parameters. Colored solid lines in (D) correspond to the modes depicted in (C).}
\end{figure*}

We turn our attention to a device (where $L = 2~\milli\meter$) operating in the single-mode regime that is well represented by the model shown in Fig.~\ref{fig:Overview}C and in equation~\eqref{eqn:total_ham}. Using coherent spectroscopy, we measure the frequency-dependent phase shift of the reflected probe, shown in Fig. \ref{EITdata}. The force generated by the pump-probe interaction in the optomechanical transducer excites the phononic wire and leads to a significant modification of the probe response. The full probe response in the weak coupling regime ($g_0\ll\kappa$) is captured by the sideband reflection coefficient
\begin{eqnarray}
r(\omega) = 1 - \cfrac{\kappa_{e}}{i(\Delta-\omega) + \frac{\kappa}{2}  +\frac{g_0^{2}\nc}{i(\omega_{m}-\omega)+ \frac{\gammai}{2} + \sum_{k} \frac{f_{k}^{2}}{i(\omega_{k}-\omega) + \frac{\gammaki}{2}}}}, \nonumber
\end{eqnarray}
derived from equation~\eqref{eqn:total_ham} (see SM). The system parameters in this expression are estimated by minimizing the difference between the measured and theoretical phase response. The same values of the parameters are used in a fit that takes into account measurements at five different powers ranging from $P = 43~\micro\watt$ to $P = 254~\micro\watt$. The results of the estimation procedure (see SM) are shown in the solid lines of Fig.\ref{EITdata}A,B. We find good agreement between the data and the model for the obtained estimates of the waveguide mode coupling rates $f_{k}$, and intrinsic losses $\gamma_{k,\textrm{i}}$. In addition, the estimated parameters are consistent with simulations and previous measurements of intrinsic mechanical losses~\cite{Patel2017}. The mean values for the estimated parameters on the measured device are $\overline{f_{k}}/2\pi = 310 \pm 67~\kilo\hertz$, $\overline{\gamma}_{k,\textrm{i}}/2\pi = 22 \pm 2~\kilo\hertz$. Due to disorder, the spacing between pairs of consecutive extended modes ($\Delta\omega_k$) varies from mode to mode. Nonetheless, the mean value of this spacing,  $\overline{\Delta{\omega_{k}}}/2\pi = 1.6 \pm 0.2~\mega\hertz$, agrees well with an expected FSR ($v_{g}/2L = 1.7~\mega\hertz$) for the $2$-mm-long phononic wire. Using the relation $f_{k} = \sqrt{\omega_{\textrm{FSR}}\gamma_{e}/2\pi}$ (see SM) we find that the extrinsic mechanical loss rate, coupling the localized resonator to the waveguide, is $\gamma_{e}/2\pi = 386~\kilo\hertz$ consistent with simulations of phonon loading in the fabricated design (see SM) and exceeding the intrinsic damping of transducer mode $\gamma_{\textrm{i}}$. 

Once the intrinsic parameters are determined, we estimate the waveguide readout strength by diagonalizing the matrix describing the dynamics of the coupled system under laser drive. The waveguide mode damping rates $\gammak > \gamma_{k,\text{i}}$ are the imaginary components of the eigenfrequencies and depend on $\nc$. 
We define waveguide mode cooperativity for mode $k$ at intracavity photon number $\nc$ as $C_{k,\textrm{wg}}|_{\nc} \equiv \frac{\gammakS|_{\nc}}{\gammak|_{\nc = 0}}$ where $\gammakS|_{\nc} \equiv \gammak|_{\nc} - \gammak|_{\nc=0}$ is the optically induced \emph{sympathetic cooling} rate of the extended modes of the phononic wire, analogous to experiments in atomic physics~\cite{Jckel2015}. A cooperativity larger than 1 represents coupling of an extended mode to the optical readout channel that is stronger than its damping. $C_{k,\textrm{wg}}$ is plotted in Fig.~\ref{EITdata}D for several $k$. We reach the large cooperativity regime, $C_{\textrm{wg}} > 1$ for several of the extended modes that are close in frequency to the transducer mode frequency. The peak cooperativity achieved is $C_{k,\textrm{wg}} \approx 6.3$ for an extended mode detuned by $\delta_{k}/2\pi \approx 630~\kilo\hertz$ from the localized transducer resonance.  In contrast to a typical optomechanical system where the cooling rate given by $\gammaOM$ increases linearly with input laser power, the extended mode cooperativity may display nonmonotonic behavior. The sympathetic cooling rate in the weak coupling limit $f_{k} \ll \gammaOM \ll \kappa$ is approximately  
\begin{equation}
\gammakS \approx \frac{|f_{k}|^{2}\gammaOM}{\deltak^{2} + \gammaOM^{2}}.
\end{equation}
At low optical power ($\gammaOM \ll \delta_k$), increasing the pump intensity causes an increase in the damping of the extended mode since the transducer mechanical response covers a broader bandwidth. At higher optical powers ($\delta_k \ll \gammaOM$), we are only reducing the total quality factor of the transducer resonance, diminishing the readout rate in a manner analogous to a reduced Purcell enhancement.

\begin{figure}[!h]
\includegraphics[scale=1.0]{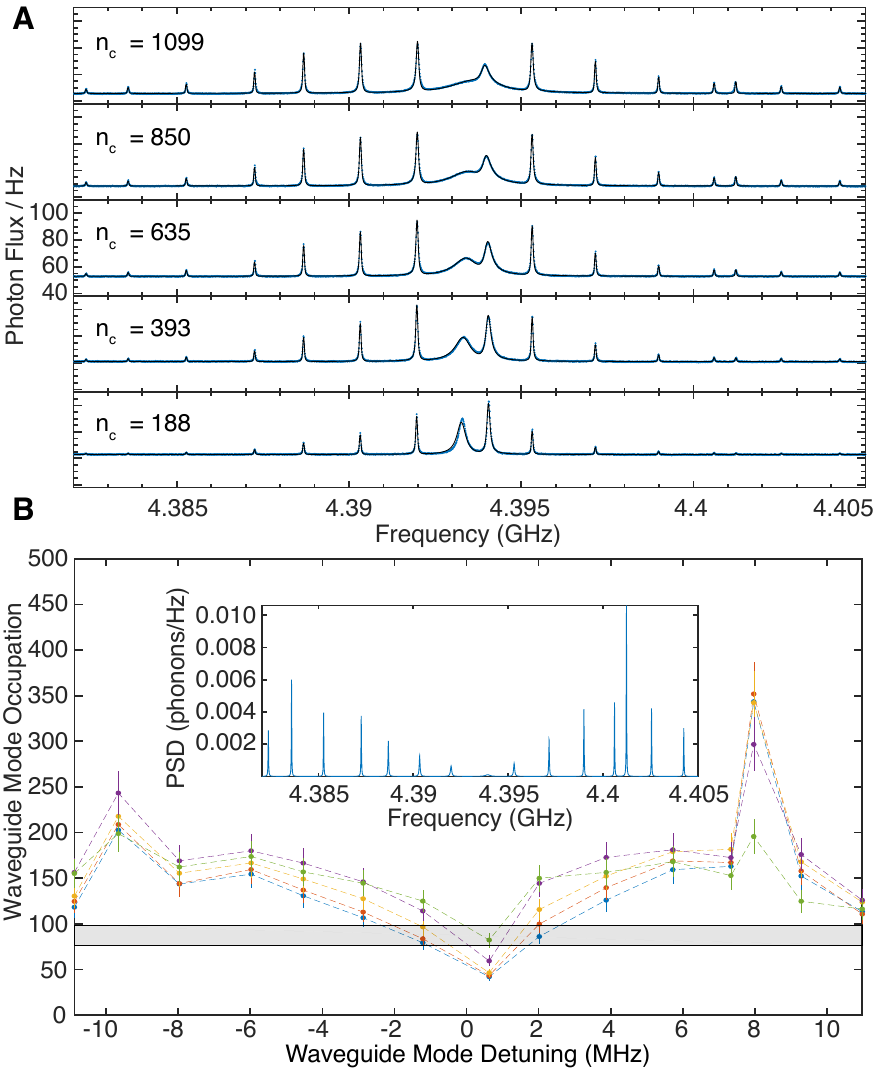}
\caption{\label{Thermaldata} \textbf{Thermal Spectroscopy and Signatures of Multimode Cooling}   \textbf{(A)} Experimental results showing the total thermal noise detected on reflection at the labeled intracavity photon numbers. The spectra are normalized by the system gain. Data points are shown in blue. The black line is a fit to theory with waveguide parameters fixed from coherent spectroscopy.\textbf{(B)} Extracted bath occupancies from the model showing cooling near resonance. The blue, red, yellow, purple and green correspond to intracavity photon numbers 1099, 850, 635, 393 and 188,respectively. The gray band shows one standard deviation uncertainty in the intrinsic phonon bath occupancy. Error bars denote 10 percent error in the system gain estimate. (Inset) Plot of the total waveguide mode power spectral density inferred from model parameters, for $n_{c} = 1099$.
Each mode is integrated to yield the occupancies plotted in (B).}
\end{figure}

The calibrated noise power spectra for the same device are shown in Fig.~\ref{Thermaldata}A. Using the same values for system parameters obtained from the fitting procedure described above, we estimate the thermal noise injected into the system by fitting to a model that includes noise inputs. The black lines in Fig.~\ref{Thermaldata}A are fits to theory, using the noise temperatures as the only free parameters (see SM). The mode occupancy for both the localized  and the extended modes are calculated from the measured noise powers and shown in Fig.~\ref{Thermaldata}B.  We observe optical cooling of the extended modes of the phononic wire. The minimum waveguide mode phonon occupancy achieved is $n_{\textrm{phon}} = 42 \pm 4$, a factor of two below the intrinsic bath occupancy of $n_{\textrm{i}} = 87\pm11$. This intrinsic bath occupancy, denoted by the gray band in Fig.~\ref{Thermaldata}B  and measured at low power across many devices~\cite{Chan2011} in the same experimental run, is smaller than the occupancy of detuned extended modes. 
For the extended mode nearly resonant with the transducer, the factor of two cooling  achieved is three times less than what would be expected naively from $C_\text{wg}\approx 6$. This is due to the presence of optical absorption induced heating of the waveguide (see SM).

In addition to optical absorption, two other nonidealities of the phononic waveguide are the persence of loss and fabrication disorder. A simple estimate for the phonon loss rate per unit length is given by $\alpha_{\textrm{m}} = \gammaki / v_{\textrm{g}} \approx 0.88~\textrm{dB} \per\centi\meter$, values comparable to silicon photonic wires~\cite{Bogaerts:05,VanLaer2015}. This loss is likely thermally activated by nonlinear phonon processes and $\alpha_{\textrm{m}}$ would be orders of magnitude smaller at lower temperatures ($10~\text{mK}$)~\cite{Meenehan2014}. For large classical signals at $10~\text{K}$, we estimate a propagation distance of $L_{3\text{dB,c}}\approx 3~\text{cm}$. At the $\approx 10~\text{K}$ temperature, the injection of thermal phonons from the bath leads to a more rapid decoherence of quantum signals of roughly $\alpha_{\textrm{m,th}} = (n_\text{i}+1)\gammaki / v_{\textrm{g}}$ and a phonon being injected into the phononic wire after  $L_{3\text{dB,q}}\approx0.5~\text{mm}$ of propagation. This decoherence would also be far smaller at $10~\text{mK}$ temperatures.

To faithfully transmit information along a wire, we must also consider the effect of disorder. Using a disordered tight-binding model of the waveguide we simulate both the single and multi-band case and find that most of the features of the thermal and coherent spectra and resulting waveguide parameters can be reproduced theoretically for levels of disorder consistent with experiment. The measured disorder in the waveguide, \emph{i.e.}, the distribution of $\Delta\omega_k$, allows us to estimate the dephasing of an itinerant excitation propagating in the phononic wire caused by random and static frequency-dependent phase shifts (see SM). This dephasing limits transmission to  $L_{3\text{dB},\delta}\approx 6~\text{mm}$.

Our demonstration of single-channel conduction of highly confined phonons is an important step in the development of phononic systems. Moreover, we have demonstrated an efficient interface between photonic and phononic degrees of freedom, and shown that optomechanical systems can be used as on-chip resources for cooling and control of phononic circuits. The techniques of quantum optomechanics provide us with a powerful toolbox for manipulating propagating phonons on the surface of a chip. In turn, the phononic guided wave structures can enhance optomechanical experiments by connecting elements spaced by millimeters, for example in mechanically mediated microwave-to-optical converters~\cite{Vainsencher2016,Safavi-Naeini2010a,Balram2017} and phonon routers~\cite{Fang2016}. Finally, the demonstrated mechanical waves have three to five orders of magnitude higher confinement than surface and bulk acoustic waves in the same frequency range, opening the way to the study of phonon-coupled emitters~\cite{Pichler2017,Lee2016,Ovartchaiyapong2014b}, two-level systems~\cite{Suh2013}, and other phonon nonlinearities.

\textbf{Acknowledgements} This work was supported by NSF ECCS-1509107, ONR MURI QOMAND, and start-up funds from Stanford University. ASN is supported by the Terman and Hellman Fellowships. RNP is supported by the NSF Graduate Research Fellowships Program. The authors thank Timothy McKenna, Alex Wollack, and Patricio Arrangoiz-Arriola for experimental assistance and Raphael Van Laer and Marek Pechal for helpful discussion. Device fabrication was performed at the Stanford Nano Shared Facilities (SNSF) and the Stanford Nanofabrication Facility (SNF). The SNSF is supported by the National Science Foundation under Grant No. ECCS-1542152. This material is based upon work supported by the National Science Foundation Graduate Research Fellowship under Grant No. DGE-1656518.

%

\pagebreak
\widetext
\begin{center}
\textbf{\large Supplemental Materials: A single-mode phononic wire}
\end{center}
\setcounter{equation}{0}
\setcounter{figure}{0}
\setcounter{table}{0}
\makeatletter
\renewcommand{\theequation}{S\arabic{equation}}
\renewcommand{\thefigure}{S\arabic{figure}}
\renewcommand{\bibnumfmt}[1]{[S#1]}
\renewcommand{\citenumfont}[1]{S#1}

\section{Materials and Methods}
\section{Definitions}
Here we provide a table of parameters referred to in the text:  

\begin{table}[H]
\caption{Definition of Parameters.} \label{tab:parameters}
\begin{center}
\begin{tabular}{ccc}
Parameter & Description & Approximate Values \\ 
\hline
$g_0/2\pi$ & single photon OM coupling rate & 700kHz - 1MHz \\
$\kappai/2\pi$ & intrinsic cavity loss rate & 700MHz \\
$\kappae/2\pi$ & extrinsic cavity loss rate & 300MHz \\
$\gammai/2\pi$ & intrinsic mechanical loss rate & 20 kHz - 100 kHz \\
$\nc$ & intracavity photon number & 100 - 1000 \\
$\omegam/2\pi$ & mechanical frequency & 4-5 GHz \\
$\omegac/2\pi$ & optical cavity frequency & 194 THz \\
$\omegaL/2\pi$ & laser frequency & 194 THz\\
$\Delta/2\pi$ & laser-cavity detuning& 4-5GHz \\
$f_{k}/2\pi$ & waveguide mode coupling rate & 300 kHz \\
$\gammaki/2\pi$ & intrinsic waveguide mode loss rate & 20 kHz \\
$\omegak/2\pi$ & waveguide mode frequency & 4-5 GHz \\
$\deltak/2\pi$ & relative waveguide mode frequency ($\omegak - \omegam$) & 0 - 10 MHz\\
$\gammakS/2\pi$ & waveguide sympathetic cooling rate & 0-250 kHz\\
$\gammak/2\pi$ & waveguide total linewidth & 20-300 kHz\\
$\Cwg$ & waveguide cooperativity &0-6\\
$\gammae/2\pi$ & waveguide coupling rate& 300kHz \\
$\omegaFSR/2\pi$ & waveguide free spectral range & 1.5MHz \\
$\vg$ & group velocity & 7000 m/s \\
$\alphaM$ & phonon loss per unit length & 1 $\textrm{dB}\per\centi\meter$\\
\end{tabular}
\end{center}
\end{table}

\section{Devices}

\subsection{Photonic Coupler Design}
\begin{figure}[h]
\includegraphics[scale=1.0]{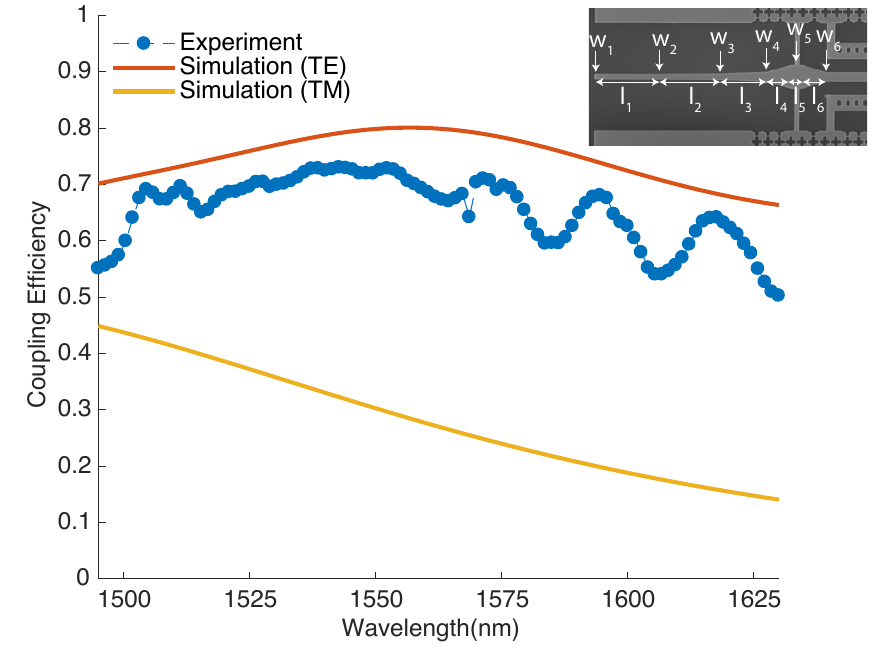} 
\caption{\label{sifig:coupleroptical} Characterization of the on-chip optical coupler.}
\end{figure}

In order to generate efficient optical driving and readout of our devices we numerically optimize an adiabatic coupler which is fabricated from thin film silicon. The coupler is designed with a Gaussian mode input of beam waist $\approx 2.5\micro\meter$, in order to couple to a lensed fiber. The simulated and measured coupler performance on devices tested at room temperature is shown in Fig. \ref{sifig:coupleroptical}. We attribute most of our reduction in efficiency to splice losses within the cryostat, as well as fiber degradation with time over multiple cooling cycles. Our structure is specified by 6 width and 6 length parameters given in Table \ref{table:coupleroptical}

\begin{table}
\caption{\label{table:coupleroptical} Optical Coupler Parameters}

\begin{tabular}{cc}
Parameter & Dimension ($\micro\meter$) \\
\hline
$w_{1}$ & 0.249 \\
$w_{2}$ & 0.251 \\
$w_{3}$ & 0.292 \\
$w_{4}$ & 0.425 \\
$w_{5}$ & 0.966 \\
$w_{6}$ & 0.400 \\
$l_{1}$ & 2.66 \\
$l_{2}$ & 2.66 \\
$l_{3}$ & 1.91 \\
$l_{4}$ & 1 \\
$l_{5}$ & 0.5 \\
$l_{6}$ & 1 \\
\end{tabular}
\end{table}

\subsection{Device Geometry and Fabrication} 

The waveguide transducer is designed by numerically optimizing a one dimensional optomechanical crystal. We aim to minimize a cost function of the form: 

\begin{equation}
C \propto g_{i}g_{j} Q_{\textrm{optical}}  
\end{equation}

where the difference between the mechanical frequencies is bounded to $|\omega_{\textrm{m},i} - \omega_{\textrm{m},j}|\approx 300~\mega\hertz$.

\begin{figure}[h]
\includegraphics[scale=1.0]{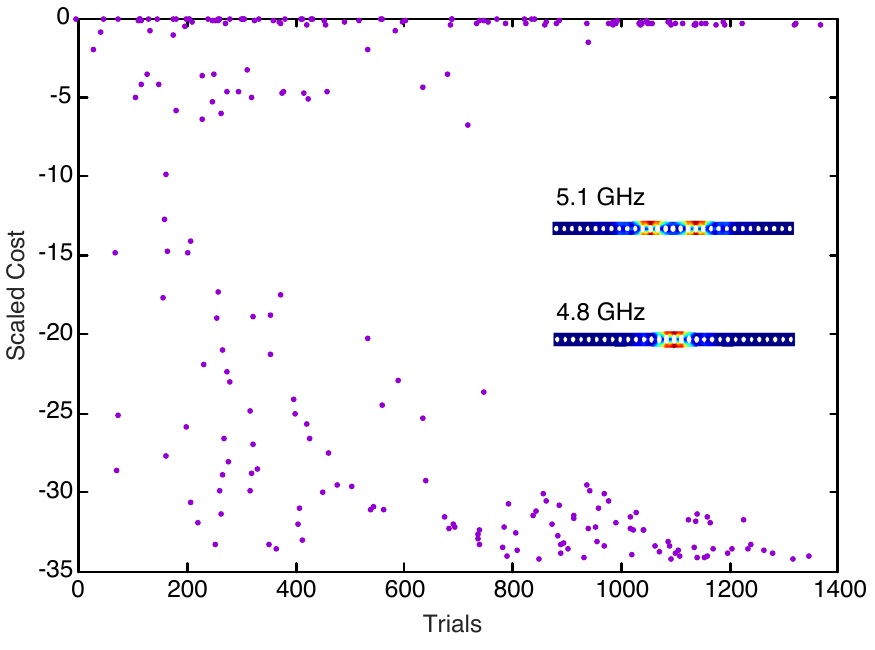} 
\caption{\label{sifig:gacost} Cost function for repeated simulations run using a genetic algorithm. The resulting mode profiles for two mechanical modes are shown in the inset.}
\end{figure}

The behavior of the cost function vs trial number from a genetic algorithm is shown in Fig.~\ref{sifig:gacost}

\begin{figure}[h]
\includegraphics[scale=1.0]{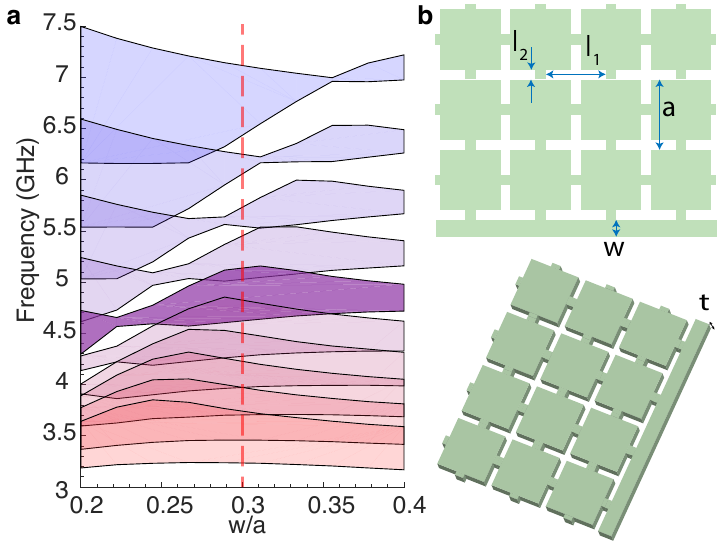}
\caption{\label{sifig:gapmap} Gapmap and geometry description. (a) A gap map showing the bandgap (shaded region) as a function of w and a. a is swept from $300~\nano\meter$ to $600~\nano\meter$. The vertical dashed line shows the value of w we use in the experiment. The dark purple region denotes $a\approx 467~\nano\meter$. $a = 450~\nano\meter$ is used in the experiment.(b) Definition of the device parameters.}
\end{figure}

The phonon waveguide geometry is shown in Fig.\ref{sifig:gapmap} and the nominal parameters are given in Table \ref{tab:phongeomparameters}. 

\begin{table}[H]
\caption{Nominal Phononic Wire Geometry Parameters} \label{tab:phongeomparameters}
\begin{center}
\begin{tabular}{ccc}
Parameter& Description & Value \\ 
\hline
$a$ & lattice constant &  $450~\nano\meter$ \\
$w$ & waveguide width & $0.3 \times a$ \\ 
$l_{1}$ & cross length & $0.9 \times a$ \\
$l_{2}$ & cross width & $0.25\times a$ \\
\end{tabular}
\end{center}
\end{table}

Our devices are fabricated using a $220~\nano\meter$ silicon on $3~\micro\meter$ oxide wafer (SOITEC) with resistivity of approximately $~15~\ohm \cdot \centi\meter$. All devices described in this work are on the same $5 \times 10~\milli\meter$ chip which is diced along one edge for optical access. The fabrication details for the cavity transducer used here are described in \cite{Patel2017}. 

We varied device waveguide lengths, coupling parameters, and lattice constants. The overall optical scaling of the transducers was also varied; devices chosen for the experiment had optical modes in the bandwidth of our EDFA (1529 - 1564 nm). 
The devices in the lattice constant sweep of Figure 2 have $L=3~\milli\meter$, with a hole offset of $\delta y = 30~\nano\meter$, coupling was design to be approximately $288~\kilo\hertz$. The offset applies uniformly to all air-holes in the direction orthogonal to both the propagation direction and the normal vector of the chip surface, as described in previous work \cite{Patel2017}. The devices shown in Figs.3 and 4 are of $L =  2~\milli\meter$ and $\delta y = 40~\nano\meter$. The phonon coupler design process will be elaborated in future work.

\section{Phonon waveguides probed through a cavity}


In this section we provide the theory and definitions of how waveguide properties can be characterized by studying the thermal and excitation spectrum of a resonator that is coupled to a terminated waveguide. This is important for phononic systems in particular. In most optical and electromagnetic experiments, excitation of guided
waves is technically simple and localized resonant modes are probed
through waveguides. With optomechanical crystals we have somewhat
the opposite situation on the mechanical side of the problem. The localized phonon mode is more simply probed by the optical field and so we are left in a
situation where we need to probe a phonon waveguide through the phonon cavity. 

\subsection{Setting up the problem}

\begin{figure*}[h]
\includegraphics[width=\textwidth]{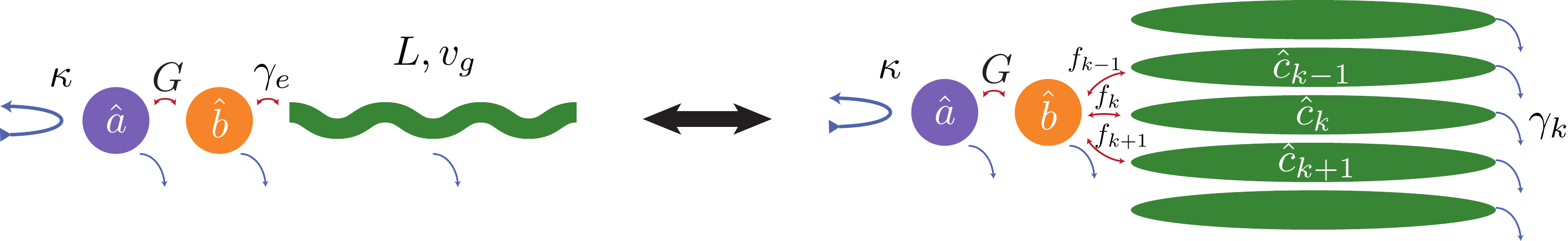}
\caption{\label{sifig:coherent_vs_thermal} Characterizing a terminated waveguide as a series of extended standing-wave modes.}
\end{figure*}

The Hamiltonian (here we take $\hbar=1$) of the waveguide, and the interaction of the waveguide
with our system can be expressed as
\begin{eqnarray}
\op H{\textrm{WG}}= &  & \sum_{k_{n}}\omega(k_{n})\opd c{k_{n}}\op c{k_{n}}\label{eq:H_WG_disc}\\
\op H{\text{sys-WG}}= &  & -i\sqrt{\frac{v_{g}\gamma_{e}}{L}}\sum_{k_{n}}\left[\op c{k_{n}}\opd b{}-\textrm{h.c.}\right]\label{eq:H_sysWG_disc}\\
\op H{\text{sys}}= &  & \omega_{m}\opd b{}\op b{}\label{eq:H_sys_disc}
\end{eqnarray}
 note that energy of the modes $\omega_{k_{n}}$ can be expressed
over the small bandwidth around the mechanical frequency that we're
interested in as $\omega_{k_{n}}=\omega_{m}+v_{g}k_{n}$. So we can
rotate out the mechanical frequency from these equations and we get
\begin{eqnarray}
\op H{\textrm{WG}}= &  & \sum_{k_{n}}v_{g}k_{n}\opd c{k_{n}}\op c{k_{n}}\label{eq:H_WG_disc-1}\\
\op H{\text{sys-WG}}= &  & -i\sqrt{\frac{v_{g}\gamma_{e}}{L}}\sum_{k_{n}}\left[\op c{k_{n}}\opd b{}-\textrm{h.c.}\right]\label{eq:H_sysWG_disc-1}\\
\op H{\text{sys}}= &  & 0\label{eq:H_sys_disc-1}
\end{eqnarray}

\subsubsection{The continuum limit}

In the case of an infinite waveguide where we can take the limit $L\rightarrow\infty$,
we are converting sums to integrals. Then
\begin{eqnarray*}
v_{g}\sum_{k_{n}}k_{n}\opd c{k_{n}}\op c{k_{n}} & = & v_{g}\sum_{k_{n}}k_{n}\opd c{k_{n}}\op c{k_{n}}(\Delta k)^{-1}\Delta k\\
 & \rightarrow & v_{g}\int dk~k\opd c{}(k)\op c{}(k).
\end{eqnarray*}
and
\begin{eqnarray*}
\sum_{k_{n}}\left[\op c{k_{n}}\opd b{}-\textrm{h.c.}\right] & = & \sum_{k_{n}}\left[\sqrt{\Delta k}\op c{}(k_{n})\opd b{}-\textrm{h.c.}\right](\Delta k)^{-1}\Delta k\\
 & = & \frac{1}{\sqrt{\Delta k}}\sum_{k_{n}}\left[\op c{}(k_{n})\opd b{}-\textrm{h.c.}\right]\Delta k\\
 & \rightarrow & \sqrt{\frac{L}{2\pi}}\int dk~\left[\opd c{}(k)]\op b{}-\textrm{h.c.}\right].
\end{eqnarray*}
with $\Delta k=2\pi/L$ and $\op c{}(k_{n})=\op c{k_{n}}/\sqrt{\Delta k}=\sqrt{\frac{L}{2\pi}}\op c{k_{n}}$.
With these substitutions we use can the continuum equations:
\begin{eqnarray}
\op H{\textrm{WG}}= & v_{g}\int dk~k\opd c{}(k)\op c{}(k)\label{eq:H_WG_cont}\\
\op H{\text{sys-WG}}= & -i\sqrt{\frac{v_{g}\gamma_{e}}{2\pi}}\int dk~\left[\opd c{}(k)]\op b{}-\textrm{h.c.}\right].\label{eq:H_WG-sys_cont}\\
\op H{\text{sys}}= & 0
\end{eqnarray}

These equations lead to damping of the mechanical mode $\op b{}$ at rate $\gamma_{\textrm{e}}$.

\subsubsection{Coupling to extended modes in a waveguide}

The interaction in equation (\ref{eq:H_sysWG_disc-1}) has the relevant
interaction rate 
\begin{align*}
f= & \sqrt{\frac{v_{g}\gamma_{e}}{L}},
\end{align*}
It's important to note that $L$ here may not be the actual length
of the waveguide. The way that $L$ came into the equations of the
continuum limit was via $\Delta k=2\pi/L$, or equivalently, $\omega_{\text{FSR}}=\omega_{n+1}-\omega_{n}=v_{g}\Delta k=2\pi v_{g}/L$.
We need this to the case for the $\gamma_{e}$ defined in these equations
to represent the actual loss rate of the cavity into the infinite
waveguide. Nonetheless, this relation allows us to define
\begin{align*}
f= & \sqrt{\frac{\omega_{\text{FSR}}\gamma_{e}}{2\pi}}.
\end{align*}
This equation behaves roughly how we expect it. For example, given
a certain waveguide and coupling geometry, as we increase the length,
$\omega_{\text{FSR}}$ reduces with the length of the waveguide, as
does the overlap of the extended modes with the main mechanical mode
\textendash{} in both cases we expect the coupling to change as $f\propto L^{-\frac{1}{2}}$. 

The operators $\op ck$ now represent the extended modes of a terminated
waveguide of finite length. This leads to a coupling Hamiltonian of
the form:

\begin{eqnarray}
\op H{\textrm{WG}}= & \sum_{k}(\omega_{k}-\omega_{\text{m}})\opd ck\op ck\label{eq:H_WG_disc-1-1}\\
\op H{\text{sys-WG}}= & -i\sum_{k}f_{k}\left[\op ck\opd b{}-\textrm{h.c.}\right]\label{eq:H_sysWG_disc-1-1}
\end{eqnarray}
We've given $f$ the index $k$ for now. In an experiment we'd expect
all of parameters $f_{k}$ to be roughly equal. 

\subsubsection{Thermal Spectroscopy: Derivation of output spectrum}

\begin{figure}[H]
\includegraphics[width=\textwidth]{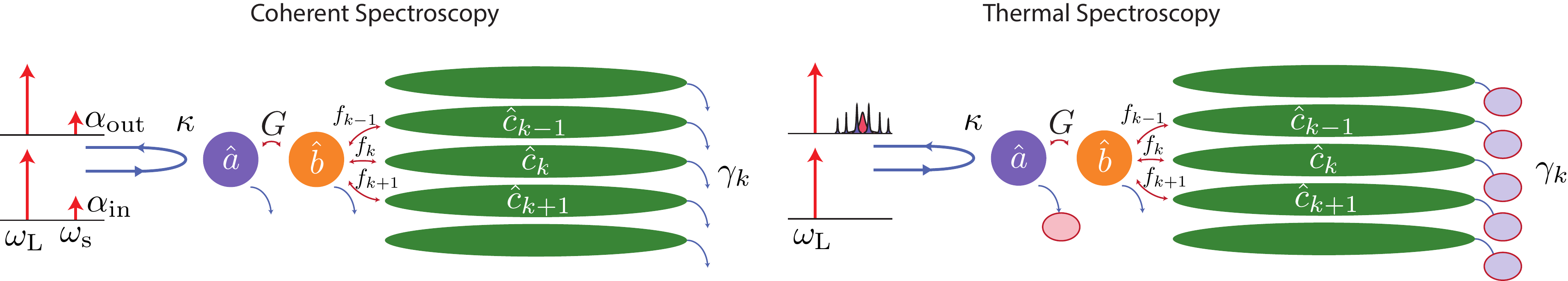}
\caption{\label{sifig:coherent_vs_thermal} Conceptual overview of coherent and thermal spectroscopy.}
\end{figure}

To get a feeling for the different parameters and possibly an understand
the important regimes of this type of coupling, we can look at the
expected spectra. There are two main types of measurement we can do.
The first is to measure the thermal spectrum out by looking at the
photons scattered via interaction with mode $\op b{}$ (this scattering
happens into the optical mode at a rate $\gamma_{\text{OM}})$. The
second is coherent spectroscopy, where we excite the system coherent
with the tone through mode $\op b{}$ and then measure the phase and
amplitude of the modified sideband. First, to set the frame we're
in, assume that our Hamiltonian is given by:
\begin{align*}
H= & \delta\opd b{}\op b{}+\sum_{k}f_{k}(\op ck\opd b{}+\text{h.c.})+\sum_{k}(\Delta_{k}+\delta)\opd ck\op ck
\end{align*}
where $\delta$ is the detuning of the drive with the mechanical mode
and $\Delta_{k}=\omega_{k}-\omega_{\text{m}}$. From here we can get
an equation that looks like
\begin{align*}
\partial_{t}\left(\begin{array}{c}
b\\
c_{1}\\
\vdots\\
c_{n}
\end{array}\right) & =\check{M}\left(\begin{array}{c}
b\\
c_{1}\\
\vdots\\
c_{n}
\end{array}\right)+\left(\begin{array}{c}
\sqrt{\gamma_{\text{OM}}}b_{\text{in}}+\sqrt{\gamma_{\text{i}}}b_{\text{in,i}}\\
\sqrt{\gamma_{\text{i},1}}c_{\text{in},1}\\
\vdots\\
\sqrt{\gamma_{\text{i},n}}c_{\text{in},n}
\end{array}\right)
\end{align*}
where

\begin{align*}
\check{M}=\left(\begin{array}{ccccc}
-i\delta-\frac{\gamma_{\text{OM}}}{2}-\frac{\gamma_{\text{i}}}{2} & -if_{1} & -if_{2} & \cdots & -if_{n}\\
-if_{1} & -i(\Delta_{1}+\delta)-\frac{\gamma_{1,\text{i}}}{2} & 0 & \cdots & \vdots\\
-if_{2} & 0 & \ddots & 0 & 0\\
\vdots & \vdots & 0 & \ddots & 0\\
-if_{n} & 0 & \ldots & 0 & -i(\Delta_{n}+\delta)-\frac{\gamma_{n,\text{i}}}{2}
\end{array}\right)
\end{align*}
We can express this as a matrix equation in the Fourier domain:

\begin{align*}
-i\omega\underline{s} & =\check{M}\underline{s}-\check{K}\underline{v}_{\text{in}}
\end{align*}
The coupling matrix is given by
\begin{align*}
\check{K}= & \left(\begin{array}{ccccc}
\sqrt{\gamma_{\text{OM}}} & \sqrt{\gamma_{\text{i}}} & 0 & \cdots & 0\\
0 & 0 & \sqrt{\gamma_{1,\text{i}}} & 0 & 0\\
0 & 0 & 0 & \ddots & 0\\
0 & 0 & \cdots & 0 & \sqrt{\gamma_{\text{i},n}}
\end{array}\right)
\end{align*}
and the input vector is $\underline{v}_{\text{in}}=\left(\begin{array}{ccccc}
b_{\text{in}} & b_{\text{in,i}} & c_{\text{in},1} & \cdots & c_{\text{in},n}\end{array}\right)^{T}$. This allows us to write the input-output relation as 
\begin{align*}
\underline{v}_{\text{out}}= & \underline{v}_{\text{in}}+\check{K}^{T}\underline{s}.
\end{align*}
 To get the output spectrum, we solve these equations, substituting
\begin{align*}
\underline{s}= & (\check{M}+\check{1}i\omega)^{-1}\check{K}\underline{v}_{\text{in}}
\end{align*}
to obtain
\begin{align*}
\underline{v}_{\text{out}}= & \left[\check{1}+\check{K}^{T}(\check{M}+\check{1}i\omega)^{-1}\check{K}\right]\underline{v}_{\text{in}}=\check{G}(\omega)\underline{v}_{\text{in}}.
\end{align*}
 Now how does this relate to the measured spectra, i.e. 
\begin{align*}
S_{\text{out}}(\omega) & =\textrm{\ensuremath{\int}d\ensuremath{\omega}}\exp(i\omega\tau)\langle b_{\text{out}}^{\dagger}(\tau)b_{\text{out}}(0)\rangle\\
\end{align*}
For that we use 
\begin{align*}
S_{\text{out}}(\omega)= & \int\textrm{d}\omega^{\prime}\langle b_{\text{out}}^{\dagger}(\omega)b_{\text{out}}(\omega^{\prime})\rangle\\
= & \int\textrm{d}\omega^{\prime}[G_{11}^{\ast}(-\omega)G_{11}(\omega^{\prime})\langle b_{\text{in}}^{\dagger}(\omega)b_{\text{in}}(\omega^{\prime})\rangle\\
 & +G_{12}^{\ast}(-\omega)G_{12}(\omega^{\prime})\langle b_{\text{in,i}}^{\dagger}(\omega)b_{\text{in,i}}(\omega^{\prime})\rangle\\
 & +G_{13}^{\ast}(-\omega)G_{13}(\omega^{\prime})\langle c_{\text{in,1}}^{\dagger}(\omega)c_{\text{in,1}}(\omega^{\prime})\rangle\\
 & \vdots\\
 & +G_{1(n+2)}^{\ast}(-\omega)G_{1(n+2)}(\omega^{\prime})\langle c_{\text{in,}n}^{\dagger}(\omega)c_{\text{in,}n}(\omega^{\prime})\rangle].
\end{align*}
Now, using the autocorrelation relations $\langle b_{\text{in,i}}^{\dagger}(\omega)b_{\text{in,i}}(\omega^{\prime})\rangle=\langle c_{\text{in,}k}^{\dagger}(\omega)c_{\text{in,}k}(\omega^{\prime})\rangle=\bar{n}\delta(\omega+\omega^{\prime})$
for the intrinsic noise channels and $\langle b_{\text{in}}^{\dagger}(\omega)b_{\text{in}}(\omega^{\prime})\rangle=0$

for the optical coupling to the mechanical mode, we obtain:
\begin{align*}
S_{\text{out}}(\omega) & =\bar{n}\sum_{j=2}^{n+2}|G_{1j}(\omega)|^{2},
\end{align*}
 i.e. the incoherent sum of the coupling through the different noise
channels.

\subsubsection{Coherent Spectroscopy: Derivation of cavity reflection coefficient }

Starting from the Heisenberg-Langevin equations describing this system
in the laser frame, with the laser red-detuned from the cavity by
roughly a mechanical frequency:

\begin{equation}
-i\omega a(\omega)=-i\Delta a-\frac{{\kappa}}{2}a-iGb-\sqrt{{\kappa_{\text{e}}}}a_{\text{in}}(\omega)
\end{equation}

\begin{equation}
-i\omega b(\omega)=-i\omega_{\text{m}}b-\frac{{\gamma_{\text{i}}}}{2}b-iGa-i{\displaystyle \sum_{k}{f_{k}c_{k}}}
\end{equation}

\begin{equation}
-i\omega c_{k}(\omega)=-i\omega_{k}c_{k}-\frac{{\gamma_{k,\textrm{i}}}}{2}c_{k}-if_{k}b
\end{equation}

and applying the optical input-output boundary condition: $a_{\text{out}}(\omega)=\sqrt{{\kappa_{\text{e}}}}a(\omega)+a_{\text{in}}(\omega)$

we get after some algebra: 

\begin{equation}
\label{sieq:reflection}
r(\omega)=\frac{{a_{\text{out}}(\omega)}}{a_{\text{in}}(\omega)}=1-\frac{{\kappa_{\text{e}}}}{i(\Delta-\omega)+\frac{{\kappa}}{2}+\frac{{G^{2}}}{i(\omega_{\text{m}}-\omega)+\frac{\gamma_{\text{i}}}{2}+\sum_{k}\frac{{f_{k}^{2}}}{i(\omega_{k}-\omega)+\frac{{\gamma_{k,\textrm{i}}}}{2}}}}
\end{equation}

Importantly, this quantity allows determination of the waveguide coupling
parameters $f_{k}$without a-priori knowledge of the temperature.
(Since the measurement is coherent, the thermal noise contributions
do not enter. The measurement bandwidth in an ideal setup can be reduced
arbitrarily close to zero). We can think of this formula as describing
the traditional optomechanical EIT effect, with additional coupling
terms to the waveguide modes. These modes, which do not couple directly
to the optics, can still be resolved in an EIT scan due to their coupling to the localized phononic cavity mode rotating at $\omega_{\text{m}}$

\section{Deducing the model from coherent spectroscopy data}
In order to determine the model parameters, it is first necessary to use equation \eqref{sieq:reflection} to predict $S_{21}(\omega)$, the measured RF scattering parameter. We start by describing the output of an ideal intensity modulator: 

\begin{equation}
\underline{\alpha}_{\textrm{out}} = 
\frac{1}{2}
\begin{pmatrix}
 1&i \\ 
 i&1 
\end{pmatrix}
\begin{pmatrix}
 1&0 \\ 
 0&e^{i\phi(t)} 
\end{pmatrix}
\begin{pmatrix}
 1&i \\ 
 i&1 
\end{pmatrix}
\underline{\alpha}_{\textrm{in}}
\end{equation}

where  $\underline{\alpha}_{\textrm{out}}  = (\alpha_{\textrm{out},1},\alpha_{\textrm{out},2})^{T}$, $\underline{\alpha}_{\textrm{in}}  = (\alpha_{\textrm{in},1},\alpha_{\textrm{in},2})^{T}$, and $\phi(t) = \phi_{\textrm{DC}} + \beta \cos (\omega t)$. In an ideal setup with the electro-optic-intensity-modulator operating at the midpoint, we take $\phi_{\textrm{DC}} = \pi/2$ The modulation strength $\beta \propto v_{\textrm{drive}}(t)$, the voltage amplitude generated by the VNA. In the rotating frame of the laser, we have $\underline{\alpha}_{\textrm{in}}  = (\alpha_{\textrm{L}},0)^{T}$ and for small $\beta$ we get for the output amplitude: 

\begin{equation}
\alpha_{\textrm{ref}}(t) = \frac{\alpha_{\textrm{L}}}{2}\Big[(1 -i) r(0) + \frac{\beta}{2}r(-\omega)e^{i\omega t} + \frac{\beta}{2}r(\omega) e ^{-i \omega t }  \Big]
\end{equation}
where we have used the reflection coefficient of Eq.\eqref{sieq:reflection} to determine the reflected amplitude $\alpha_{\textrm{ref}}(t)$. Here we are working in the case where the pump detuning is $\Delta \approx \omegam$. 
Upon detection, the VNA sees the following (AC coupled) signal: 

\begin{equation}
|\alpha_{\textrm{ref}}(t)|^{2} = \exp{(-i \omega t)}\frac{\alpha^{2}_{\textrm{L}}}{4}\Big[ \frac{\beta}{2}(1+i) r^{*}(0)r(\omega) + \frac{\beta}{2}(1-i)r(0)r^{*}(-\omega) \Big] + c.c. 
\end{equation}
The expression in brackets is proportional to the RF scattering parameter. We divide this by the far-detuned $\Delta \gg \kappa$ response to obtain an expression for the normalized scattering parameter, which does not depend on the modulation strength or the frequency response of circuit elements not related to the optomechanical system response (e.g. cable losses). This normalization yields, noting that $r(\omega) \approx 1$ for $\Delta \gg \kappa$: 

\begin{equation}
S_{21}(\omega) =  \frac{1}{2}\left[(1+i) r^{*}(0)r(\omega) + (1-i)r(0)r^{*}(-\omega)\right]
\end{equation}

Having obtained an expression for the scattering parameter, we fit the phase response by minimizing the following cost function over the system waveguide parameters $\underline{x}_{\textrm{wg}}$ and localized optomechanical parameters $\underline{x}_{\textrm{local}}$. 

\begin{equation}
f(\underline{x}_{\textrm{wg}},\underline{x}_{\textrm{local}}) = \sum_{i}\sum_{j} \Big[\angle S_{21}(\underline{x}_{\textrm{wg}},\underline{x}_{\textrm{local}},n_{\textrm{c}i},\omega_{j})  - \angle S_{21,\textrm{data}}(n_{\textrm{c}i},\omega_{j}) \Big]^{2} 
\end{equation}
where $n_{\textrm{c}i}$ denotes the $i^{th}$ intracavity photon number. The $\angle$ symbol denotes the unwrapped phase of the scattering parameter. We used five datasets corresponding to the curves shown in the main text. A summary of the parameters is given in table \ref{tab:intrinsicparameters}: 

\begin{table}[h]
\caption{Summary of Fit Results} \label{tab:intrinsicparameters}
\begin{center}
\begin{tabular}{cc}
Parameter & Value \\ 
\hline
$\gammai /2\pi$  &  $122~\kilo\hertz$ \\
$g_{0}/2\pi$  & $690~\kilo\hertz$\\ 
$\kappa /2\pi$ & $1~\giga\hertz$ \\
$\kappae/2\pi$ & $200~\mega\hertz$ \\
$\omegac/2\pi$ & $ 192.2 ~\tera\hertz$ \\
$\omegam/2\pi$ & $ 4.393 ~\giga\hertz$ \\
$\overline{\Delta}/2\pi$ & $ 4.217 ~\giga\hertz$ \\
$\overline{f_{k}}/2\pi$ & $ 310 ~\kilo\hertz$ \\
$\overline{\gammaki}/2\pi$ & $ 22 ~\kilo\hertz$ \\
$\overline{\Delta \omegak}/2\pi$ & $ 1.6~\mega\hertz$ \\
\end{tabular}
\end{center}
\end{table}

Once the intrinsic parameters have been determined, we construct the dynamical matrix from the Heisenberg-Langevin Equations as done in the previous section. We get a matrix for the complete system, working in the rotating frame of the localized mode, which looks like: 

\begin{equation}
\label{sieq:intrinsicmatrix}
\check{M}_{\textrm{total}} = 
\left[\begin{array}{ccccc}
i\Delta+\frac{{\kappa}}{2} & iG & 0 & \cdots & 0\\
iG & \frac{{\gamma_{\textrm{{i}}}}}{2} & if_{1} & \cdots & if_{k}\\
0 & if_{1} & i\delta_{1}+\frac{{\gamma_{1\textrm{{i}}}}}{2} & \cdots & 0\\
\vdots & \vdots & \vdots & \ddots & \vdots\\
0 & if_{k} & 0 & \cdots & i\delta_{k}+\frac{{\gamma_{k\textrm{{i}}}}}{2}
\end{array}\right]
\end{equation}

where we note that our choice of frame implies $\Delta \rightarrow \Delta - \omegam$, and we define $\deltak = \omegak - \omegam$. The only laser power dependent quantity here is $G$. We diagonalize ~\eqref{sieq:intrinsicmatrix} to give the modified frequencies and damping rates of the coupled system. For   the $N_{\textrm{wg}}$ eigenvalues $\{\lambda_{1} ... \lambda_{k}\}$, we can compute these total rates as: $\omega_{k,\textrm{total}} = \Im[\lambda_{k}]$ and $\gamma_{k} = 2\Re[\lambda_{k}]$. We note that $\gammai \neq 0$ when $G = 0$, since the total damping rate $\gamma_{k} \neq \gamma_{k,\textrm{i}}$. Physically, we expect this to be true for waveguide modes that are near resonance, that is, those for $\delta_{k} \ll \gammaOM$. This is also the basis for the excess heating described later and shown in Fig.\ref{sifig:heating} b. In analogy to the standard optomechanical broadening in an uncoupled system, $\gamma = \gammai + \gammaOM$ where $\gammaOM$  denotes the system measurement rate. We define $\gamma_{k} = \gamma_{k,\textrm{i}} + \gamma_{k,S}$ where $\gamma_{k,S}$ is the sympathetic cooling rate arising from coupling with the localized mode. 

\section{Deducing the thermal occupations from the coherent spectroscopy model and calibrated thermal spectrum data}

We perform optomechanical thermometry on a non-waveguide-coupled device to determine our system gain, relating the optical photon flux exiting the cavity to the RF power we detect on our electronic detector. As described in the main text, this system gain is used to normalize the thermal data measured from our waveguide-coupled device. By using the best-fit theoretical model to fit this renormalized spectrum, we can determine the individual mode occupancies. We note that we only have direct experimental access to the sum of the noise from all modes, so such an approach is necessary to perform waveguide thermometry. 

\begin{figure}[h]
\includegraphics[scale=1.0]{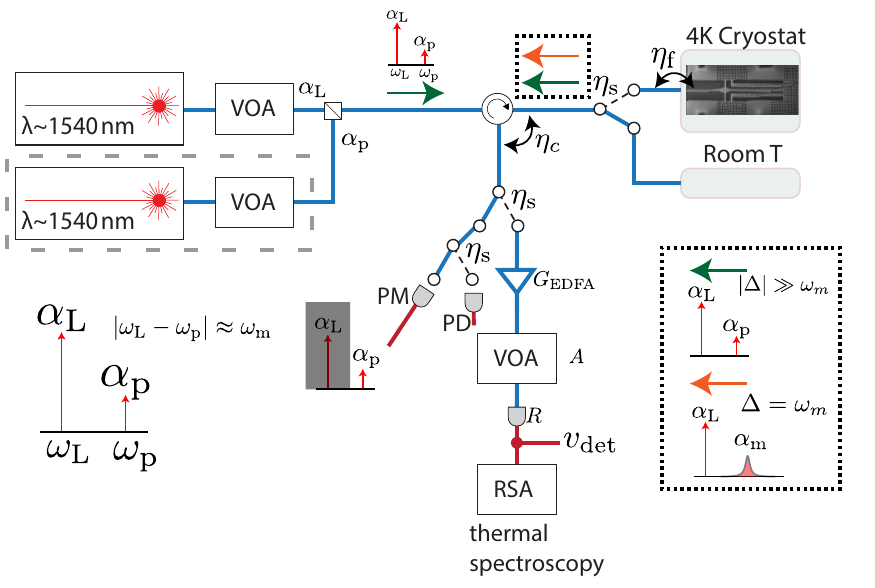}
\caption{\label{sifig:thermometry} Two-tone calibration routine for optomechanical thermometry. The green-arrow denotes the off-resonant calibration tones that are sent into the system. The orange-arrow denotes a traditional measurement scenario, in which a pump with detuning $\Delta = \omegam$ is used to readout mechanical motion. During calibration, the main laser tone can be switched off, and the probe tone power can be measured directly on a power meter, at the point marked PM. Measuring the integrated power of the beat-tone of the two lasers and dividing by probe-only optical power gives us a measure of the system gain. VOA: Variable Optical Attenuator, EDFA: Erbium Doped Fiber Amplifier, RSA: Realtime Spectrum Analyzer}
\end{figure}

A simplified diagram of our detection chain is shown in Fig.\ref{sifig:thermometry}. The idea is to use two lasers, one strong tone, and one weak probe. If the two tones are allowed to reflect off of the system (detuned from the cavity resonance), then we can measure the resulting beat signal on a high speed photodiode. Crucially, the RF power inferred from this beat signal can be compared to the flux of photons emitted by the probe laser alone, giving us a direct way to calibrate the system gain, without detailed knowledge of each individual element in the path, (such as, for example, the losses in the RF coaxial cable connecting the detector to the real-time-spectrum-analyzer). 

Let the lasers be described by $\alphaL(t) = \alphaL e^{-i\omegaL t}$ and $\alphap (t) = \alphap e^{-i \omegap t} $. Upon reflection from the device, the amplitude of the field incident on the high speed detector is: 

\begin{equation}
\alpha(t) = \sqrt{\etas^{3}\etaf^{2}\etac^{2} G_{\textrm{EDFA}} A}(\alphaL(t) + \alphap(t))
\end{equation}
where $\etas$, $\etaf$, $\etac$ denote the switch, fiber, and circulator efficiencies respectively. $G_{\textrm{EDFA}}$ denotes the Erbium-doped-fiber-amplifier (EDFA) gain, and $A$ the attenuation setting. 

The time varying part of the detected voltage is given by:

\begin{equation}
v_{\textrm{det}}(t) = R\hbar\omegap|\alpha(t)|^{2} = R\hbar\omegap \etas^{3}\etaf^{2}\etac^{2} G_{\textrm{EDFA}} A |\alphaL| |\alphap| \cos (\Delta_{\textrm{p}} t)
\end{equation}

where we have defined $R$ as the detector responsivity (which contains RF losses), $\Delta_{\textrm{p}} = \omegaL - \omegap$, and the laser frequencies are set such that $\delta \approx \omegam$ 

The voltage spectral density at the input to the spectrum analyzer is given by:

\begin{equation}
S_{vv}(\omega) = \int \langle{\vdet(t)\vdet(0)}\rangle e^{i\omega t} dt =\Big( \delta(\omega - \Delta_{\textrm{p}})+ \delta(\omega + \Delta_{\textrm{p}}) \Big)  \Big(\pi R\hbar\omegap \etas^{3}\etaf^{2}\etac^{2} G_{\textrm{EDFA}} A |\alphaL| |\alphap| \Big) ^{2}
\end{equation}

which yields the total integrated RF power: 
\begin{equation}
P_{\textrm{RF}} =\frac{1}{Z_{o}} \int_{-\infty}^{\infty} S_{vv}(\omega)d\omega =   \frac{2}{Z_{o}}\Big[\pi R\hbar\omegap \etas^{3}\etaf^{2}\etac^{2} G_{\textrm{EDFA}} A |\alphaL| |\alphap| \Big] ^{2}
\end{equation}

where $Z_{o}$ is the spectrum analyzer input impedance, typically taken as $50~\ohm$ though its value is not necessary for our calibration procedure.
The calibration is obtained by comparing this to the reflected power of the probe signal alone. This is possible because the main laser tone can be switched off. The probe signal power is: 

\begin{equation}
P_{\textrm{probe}} = \hbar\omegap \etas^{4} \etaf^{2}\etac^{2}  |\alphap|^{2}
\end{equation}

We now arrive at the expression for the system gain: 

\begin{equation}
G_{\textrm{R}} = \frac{P_{\textrm{RF}}}{P_{\textrm{probe}}} =  \frac{2\hbar\omegap}{Z_{o}}\Big[\pi R \etas\etaf\etac G_{\textrm{EDFA}} A |\alphaL| \Big] ^{2}
\end{equation}
In any given measurement run, since one has access to $P_{\textrm{RF}}$ and $P_{\textrm{probe}}$ directly, independent measurements of the parameters on the right side of this equation are not necessary.  We define a normalized gain to account for fluctuations in A, the attenuation, which is set by the user to avoid saturating the high speed detector. This normalized gain allows us to compute $G_{\textrm{R}}$ across different measurement runs since in our setup the reflected power and the product $G_{\textrm{EDFA}} A$ can be readily measured. 
\begin{equation}
G_{\textrm{R,norm}} = \frac{G_{\textrm{R}}}{(G_{\textrm{EDFA}} A)^{2}}
\end{equation}

\begin{figure}[h]
\includegraphics[scale=1.0]{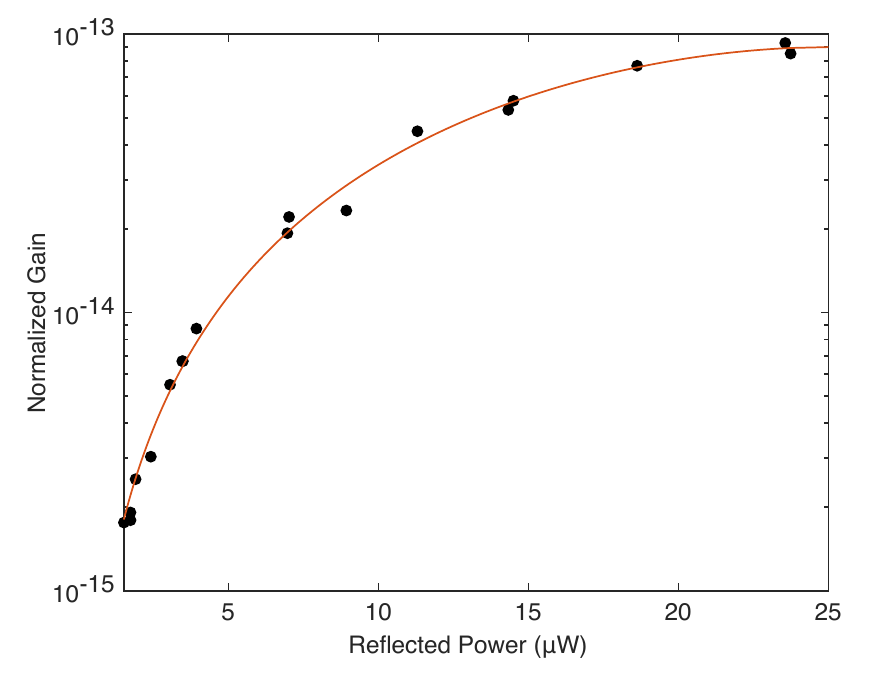}
\caption{\label{sifig:gain} Normalized gain. The black data points are gains measured from running a calibration routine in our setup with a two-tone laser drive. The red line is a third-order polynomial fit which we use to infer the system gain on other devices. We note that the normalized system gain is independent of device, since the measurements are taken at large detuning from the cavity. The reflected power refers to the laser power measured at the point PM in Fig.\ref{sifig:thermometry}}
%
%
\end{figure}

A plot of the normalized gain is shown in Fig.\ref{sifig:gain}. 

Once $G_{\textrm{R}}$ has been determined, one can calculate the photon flux in the mechanical sideband, and the resulting phonon occupancy. To illustrate this in the simplest case, we consider $\Delta = \omegam$ where we have $H_{\textrm{int}} = G (\aopd \bop + \aop \bopd)$. Then we can write a classical equation for the optics in the steady-state: 

\begin{equation}
\dot{a} = 0 =  -\frac{\kappa}{2}a - i G b 
\end{equation}

This, in addition to the input-output boundary condition $\aout = \sqrt{\kappae}a + \ain$ gives us: 

\begin{equation}
\label{sieq:outputflux}
|\alpha_{\textrm{m}}|^{2} = |\aout|^{2} = \frac{\kappae}{\kappa}\frac{4 G^{2}}{\kappa} n_{\textrm{phon}} = \frac{\kappae}{\kappa} \gamma_{\textrm{OM}} n_{\textrm{phon}}
\end{equation}
We see that this expression for the sideband photon flux (see the box in Fig. \ref{sifig:thermometry} for the $\Delta = \omegam$ case) is proportional to the measurement rate $\gamma_{\textrm{OM}}$. The RF power generated by this signal upon detection is: 

\begin{equation}
\label{sieq:RFoutput}
P_{\textrm{RF}} = \hbar\omegac |\alpha_{\textrm{m}}|^{2} G_{\textrm{R}}\etac \etaf \etas^{3}
\end{equation}

\begin{figure}[h]
\includegraphics[scale=1.0]{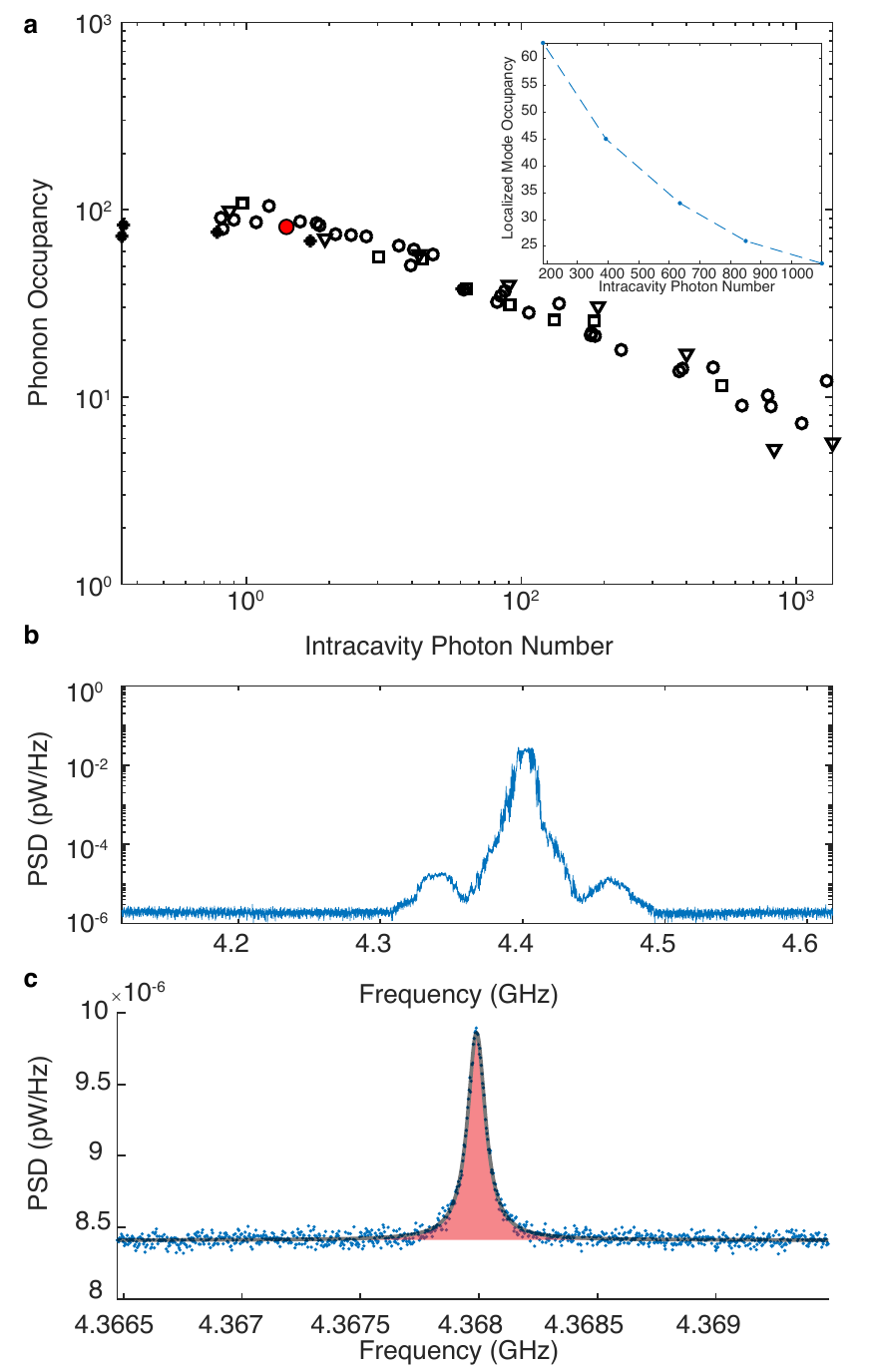}
\caption{\label{sifig:occupancies} Results of calibrated noise thermometry. (a) Mechanical thermal occupancies using the gain calibration technique described in the text. At low powers, the optomechanical back-action cooling is negligible, and the occupancies approach equilibrium with the intrinsic thermal bath. The symbols denote different devices, all on the same chip and measured on different days. The inset shows the inferred occupancy of the localized mode in a waveguide coupled device, as described in the main text. (b) Power spectral density of the two-tone beat note with $\Delta \gg \omegam$, from which an RF power can be calculated through integration. The beat-note was recording during a measurement of the point highlighted in red in part (a). (c) Power spectral density of the thermally driven mechanical mode where $\Delta \approx \omegam$. The area in red denotes the background subtracted area and is proportional to the phonon occupancy. The calibration procedure fixes this proportionality constant.}
\end{figure}

For an example of $P_{\textrm{RF}}$ see the red shaded region of Fig.\ref{sifig:occupancies}c. 
Combining \eqref{sieq:outputflux} and \eqref{sieq:RFoutput} we get an equation for the phonon occupancy: 

\begin{equation}
n_{\textrm{phon}} =\frac{1}{G_{\textrm{R}}} \Big(\frac{P_{\textrm{RF}}}{\hbar\omegac}\Big) \Big( \frac{\kappa}{\kappae \gamma_{\textrm{OM}}\etac \etaf \etas^{3}} \Big)
\end{equation}

In order to determine the thermal output spectra from our model we start with the input-output equation derived earlier: 

\begin{equation}
\label{sieq:thermalmodel}
\underline{v}_{\text{out}}= [\check{1}+\check{K}_{\textrm{out}}(\check{M}+\check{1}i\omega)^{-1}\check{K}_{\textrm{in}}]\underline{v}_{\text{in}}=\check{G}(\omega)\underline{v}_{\text{in}}.
\end{equation}

where we have defined the matrices $\Kin$ and $\Kout$ as: 

\begin{equation}
\Kin = 
\left(\begin{array}{ccccc}
\sqrt{\kappa_{\textrm{e}}} & 0 & 0 & 0 & 0\\
0 & \sqrt{\gamma_{\textrm{i}}} & 0 & \cdots & 0\\
0 & 0 & \sqrt{\gamma_{1,\textrm{i }}} & 0 & \vdots\\
0 & \vdots & 0 & \ddots & 0\\
0 & 0 & \cdots & 0 & \sqrt{\gamma_{k,\textrm{i}}}
\end{array}\right)
\end{equation}

and

\begin{equation}
\Kout = 
\left(\begin{array}{cccc}
\sqrt{\kappa_{e}} & 0 & \cdots & 0\\
0 & 0 & \cdots & 0\\
\vdots & \vdots & \ddots & \vdots\\
0 & 0 & \cdots & 0
\end{array}\right)
\end{equation}

$\Kout$ includes coupling to the optical channel only, since our experiment measures an optical output signal. From here, we can calculate the spectral density of the photon flux emitted from the system, $\aout$, as: 

\begin{equation}
S_{\aout \aout}(\omega) = \int^{\infty}_{-\infty} \langle a^{\dagger}_{\textrm{out}}(\omega)a_{\textrm{out}}(\omega^{\prime})\rangle d\omega^{\prime} = \sum^{N_{\textrm{wg}}+1}_{i=1} n_{i} |\check{G}(\omega)_{1,i+1}|^{2}
\end{equation}
where the $n_{i}$ are unknown bath occupancies that are determined by a fit to the total noise output. We define $n_{i}$ for $i=1$ to denote the localized mode input bath occupancy. In order to determine the $n_{i}$ we first normalize the experimental data,$S_{vv}(\omega)$, using the gain factor calculated in the previous section. This gives us: 

\begin{equation}
\tilde{S}_{\aout \aout}(\omega)  =  \frac{S_{vv}(\omega)}{G_{\textrm{R}} \etac \etaf \etas^{3} \hbar \omegac}
\end{equation}

We now collect the relevant part of the model output from \eqref{sieq:thermalmodel} in a matrix, with each column corresponding to the model evaluated over a range of frequencies described in the vector $\underline{\omega}$: 
\begin{equation}
\check{H}(\underline{\omega}) = 
\left(\begin{array}{cccc}
\vdots &  & \vdots & 1\\
|\check{G}_{1,2}(\underline{\omega})|^{2} & \cdots & |\check{G}_{1,N_{\textrm{wg}}+2}(\underline{\omega})|^{2} & \vdots\\
\vdots &  & \vdots & 1
\end{array}\right)
\end{equation}

The last column is added to incorporate a constant noise floor. Defining $\underline{N}$ as the vector of unknown bath occupancies, $(n_{1} ... n_{N_{\textrm{WG} + 1}},c)^{T}$ and $\underline{S}(\underline{\omega}) = \tilde{S}_{\aout \aout}(\underline{\omega})$as the experimentally measured power spectrum, we try solving for $\underline{N}$
 
\begin{equation}
\check{H} \underline{N} = \underline{S}
\end{equation}

The solution is a standard least-squares problem, and is given by: 

\begin{equation}
\underline{N} = (\check{H}^{T} \check{H})^{-1}  \check{H}^{T}\underline{S} 
\end{equation}

Finally we need to determine the occupancies of the $N_{\textrm{wg}} + 1$ mechanical modes. To do so we define the matrix (implicit in Eq.\eqref{sieq:thermalmodel}):  

\begin{equation}
G_{\textrm{int}}(\omega) = (\check{M}+\check{1}i\omega)^{-1}\check{K}_{\textrm{in}}
\end{equation}

From here we can compute the spectra for the localized mode using: 

\begin{equation}
S_{bb}(\omega) = \sum^{N_{\textrm{wg}+1}}_{i=1}  n_{\textrm{eff},i} |G_{\textrm{int},2,i+1}(\omega)|^{2}
\end{equation}
 
and similarly for the waveguide modes using: 

\begin{equation}
S_{c_{k}c_{k}}(\omega) = \sum^{N_{\textrm{wg}+1}}_{i=1}  n_{\textrm{eff},i} |G_{\textrm{int},2+k,i+1}(\omega)|^{2}
\end{equation}

where the waveguide index $k$ runs from $1$ to $N_{\textrm{wg}}$. 

From these spectra, the mode occupancies are obtained via:  

\begin{equation}
n_{b} = \frac{1}{2\pi}\int S_{bb}(\omega) d\omega
\end{equation}

\subsection{Estimation of Heating Rates}
We start with detailed balance in an optomechanical system: 

\begin{equation}
\label{ratequation}
\dot{n}_{\textrm{b}} = 0 = - (\gammaOM + \gammai)\nb + \gammai \nintrinsic + \gammap \np 
\end{equation}
where $\nb$ denotes the phonon occupancy of the mechanical mode,$\nintrinsic$ denotes the occupancy of the thermal bath at zero optical power and $\np$ denotes the occupation of the photon-induced-heating bath. We make the assumption that the rate $\gamma_{p}\ll\gammai$. This assumption is experimentally verified as we observe no additional broadening in mechanical line-width, after taking the back-action induced line-width into account.  
The first term on the right side of Eq. \eqref{ratequation} represents the rate at which phonons leave the system.
The second and third terms represent the rate at which noise enters the system, which we denote as: 

\begin{equation}
\Gamma_{\textrm{noise,b}} = \gammai \nintrinsic + \gammap \np 
\end{equation}

where, at zero optical power: 

\begin{equation}
\Gamma_{\textrm{noise,b}}|_{\nc = 0} = \gammai \nintrinsic
\end{equation}

Using these relations, we can define a heating rate: 
\begin{equation}
\label{heatinglocal}
\Gamma_{\textrm{heating,b}} = \Gamma_{\textrm{noise,b}} -\Gamma_{\textrm{noise,b}}|_{\nc=0}= \Gamma_{\textrm{noise,b}} -\gammai  n_{\textrm{i}} 
\end{equation}

For each waveguide mode $\ckop$ we have a similar relation:
\begin{equation}
\label{heatingwg}
\Gamma_{\textrm{heating,}k} = \Gamma_{\textrm{noise,}k} -\gammaki  n_{\textrm{i}} 
\end{equation}

\begin{figure}[H]
\includegraphics[width=\textwidth]{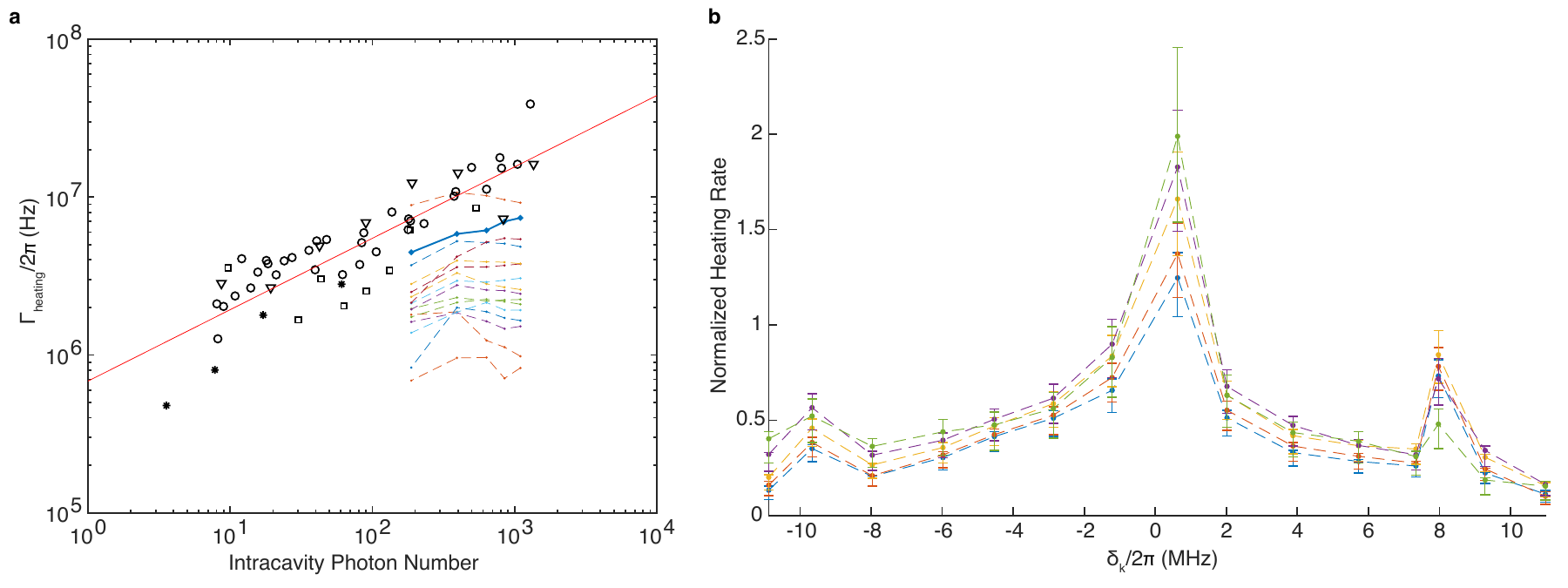}
\caption{\label{sifig:heating} Heating rates estimated from thermomechanical noise spectra. (a) Heating rates plotted for non-waveguide coupled devices (black data points, different shapes represent different trials) and a waveguide coupled device (colored data points). The solid blue line shows the localized mode heating rate. The solid red line is a linear fit, giving $\Gamma_{\textrm{heating}} \propto n_{\textrm{cav}}^{0.45}$.(b) Waveguide mode heating rates normalized by the localized mode heating rate, plotted verses waveguide detuning from the localized mechanical mode. A resonant heating effect is observed, which we attribute to line-width broadening of resonant waveguide modes at zero power. Error bars derived from one standard deviation uncertainty in the intrinsic thermal occupancy.}
\end{figure}

In the limit of low laser power where $\gammaOM\approx0$ and $\np\approx0$, we can estimate the intrinsic occupation from Eq. \eqref{ratequation} ($\nintrinsic\approx\nb$). Since we have independent measurements of $\gammai$ and $\gammaki$, and best-fit estimates of the noise rates as described in the previous section (setting $\Gamma_{\textrm{noise}} \approx \gammai n_{\textrm{eff}} $), we can directly determine the heating rates in Eqs.\eqref{heatinglocal} - \eqref{heatingwg}. The results are shown in Fig. \ref{sifig:heating}

\section{Disordered Waveguides}
Another way of modeling the waveguide is to spatially discretize it into small pieces, with each piece considered as an optical cavity. The effect of disorder could be described by introducing randomness to the on-site frequency, while keeping photon hopping rate between nearest neighbor sites as a constant $J$. Therefore the Hamiltonian for this randomized tight-binding model is:
\begin{equation}
H = \sum_{j} \omega_j a_j^\dagger a_j + J \sum_{j} (a_j^\dagger a_{j-1} + \text{h.c.})
\end{equation}
where the on-site frequency $\omega_j$ is a Gaussian random variable with mean value $\Omega$ and variation $\delta \omega$.

The model parameters $\Omega$ and $J$ could be calculated using the dispersion relation of the waveguide. Assuming the lattice constant of the discretization is $a$ and the waveguide has no disorder, i.e., $\delta \omega = 0$, then the dispersion relation of this tight binding model is
\begin{equation}
E = \Omega + 2J \cos (ka)
\end{equation}
where $k$ is the wavevector. Without loss of generality, we only consider the band structure near $ka=\pi/2$. Therefore the hopping term $J$ and the group velocity $v_g$ are related by $v_g = 2Ja$ and $\Omega$ could be approximately taken as the central frequency of the bandgap.

\begin{figure*}[h]
\includegraphics[scale=1.0]{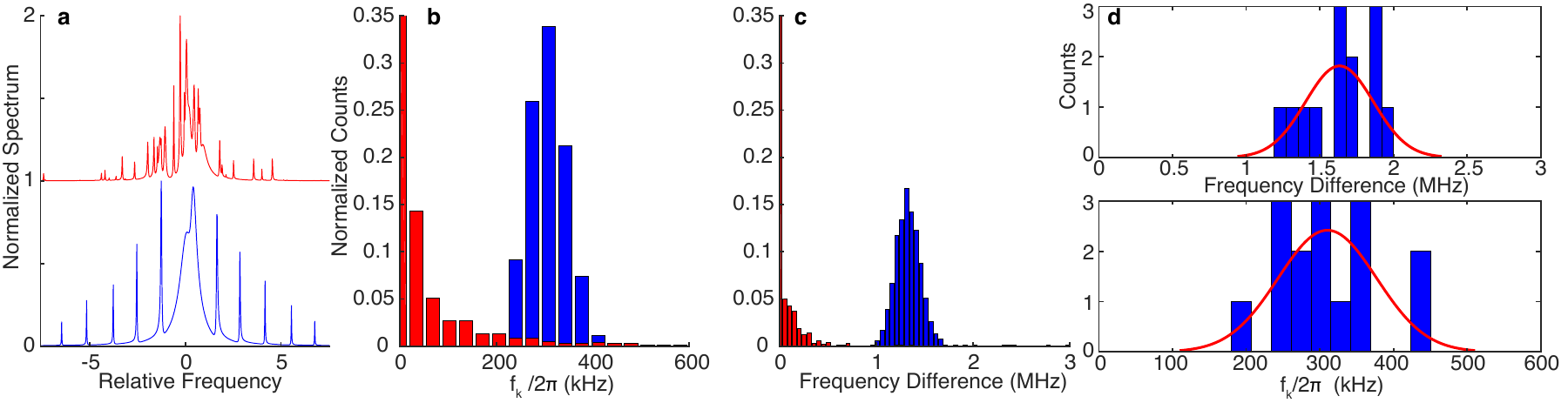} 
\caption{\label{sifig:disorder} Tight-binding Model Diagonalization Results. (a) For a disorder parameter $\delta\omega/2\pi = 10~\mega\hertz$ we obtain simulated thermal spectra for the multi-band (red) and single band (blue) case. The spectra are normalized so they range from 0 to 1. The x-axis units are multiples of the mechanical linewidth. (b) Histogram showing the distribution of coupling rates obtained after diagonalizing the tight-binding model. The blue bars show the single-mode case with standard deviation $\sigma = 40~\kilo\hertz$. (c) Histogram showing the corresponding frequency difference distribution. The blue bars show the single-mode case with standard deviation $\sigma = 126~\kilo\hertz$. (d) Histograms extracted from the linear spectroscopy model fit to the experimental data. The upper (lower) distributions have standard deviations $\sigma = 231~\kilo\hertz$ ($\sigma = 67~\kilo\hertz$)}
\end{figure*}

The variation $\delta \omega$ of the on-site frequency comes from imperfection of fabrication. A rough estimate of its values can be obtained by considering the distribution of mechanical frequencies obtained from fabricating many nominally identical mechanical resonators with localized modes and considering the statistics of the distribution of the localized mode frequencies. The disorder in these frequencies is measured to be about $8~\text{MHz}$. We chose $\delta\omega/2\pi = 10~\mega\hertz$ to be the on-site frequency disorder and simulated the tight-binding model with disorder.  The results of these simulations and a comparison to experimental data are shown in Fig.\ref{sifig:disorder} a-d. Notably we find that the standard deviation in $\Delta\omega$ is about $200~\text{kHz}$, nearly a factor of 50 smaller than the $10~\text{MHz}$ on-site frequency disorder in the tight binding model. In effect the extended modes sense disorder averaged over many unit cells and so we expect the fluctuations to diminish by roughly $\sqrt{a/L}$, i.e., a factor of $50$ for a 2 to 3 millimeter long waveguide.

\subsection{Dephasing due to Frequency Disorder}

We start with a wavefunction describing the initial state of the waveguide expanded in an orthonormal basis:

\[
\varphi(x,t=0)=\frac{1}{\sqrt{L}}\sum_{k}c_{k}e^{-ikp_{0}x}
\]
where the coefficients $c_{k}$ satisfy $\sum_{k}|c_{k}|^{2}=1$ and we define a wave number $p_{0}=\frac{2{\omega_{FSR}}}{v_{g}}$. 

For $t>0$ in the undisordered case we have:

\[
\varphi(x,t)=\sum_{k}c_{k}e^{-ikp_{0}x}e^{-ik\omega_{k}t}
\]
In the case of frequency disorder: 
\[
\varphi_{d}(x,t)=\sum_{k}c_{k}e^{-ikp_{0}x}e^{-i(k\omega_{k}+\delta\omega_{k})t}
\]
where $\delta\omega_{k}$ is a zero mean normal random variable with variance $\sigma^{2}$.

We define pulse fidelity as the quantity:

\[
F(t)=\langle\frac{1}{L}\int_{0}^{L}dx\varphi_{d}^{*}(x,t)\varphi(x,t)\rangle
\]

This gives: 

\[
F(t)=\langle\sum_{kk^{'}}c_{k'}^{*}c_{k}\frac{1}{L}\int_{0}^{L}dxe^{i(k^{'}-k)p_{0}x}e^{i[(k^{'}-k)\omega_{k}+\delta\omega_{k^{'}}]t}\rangle
\]

\[
=\langle\sum_{kk^{'}}c_{k'}^{*}c_{k}\delta_{kk^{'}}e^{i[(k^{'}-k)\omega_{k}+\delta\omega_{k^{'}}]t}\rangle
\]

\[
=\langle\sum_{k}|c_{k}|^{2}e^{i\delta\omega_{k}t}\rangle 
=\langle e^{i\delta\omega_{k}t}\rangle=e^{-\frac{t^{2}\sigma^{2}}{2}}
\]

To calculate $L_{3\text{dB},\delta}$ we set $F(t)=1/2$.

\end{document}